\documentclass[12pt,english,floatfix,nofootinbib,superscriptaddress,aps,prd,preprint]{revtex4-2}
\usepackage[utf8]{inputenc}       
\usepackage[english]{babel}       
\usepackage[utf8]{inputenc}
\usepackage{textcomp}
\usepackage{amsmath,amsthm,amssymb,amsfonts,mathrsfs,amsbsy} 
\usepackage{tensor}               
\usepackage{slashed}  
\usepackage{bbm}
\usepackage{esint}    
\usepackage[a4paper, margin=1.6cm]{geometry}
\usepackage{cancel}    
\usepackage{caption}
\usepackage{booktabs}
\usepackage{graphicx}             
\usepackage{natbib}
\usepackage{float}                
\usepackage{subfig}               
\usepackage[font=small,labelfont=bf]{caption} 
\usepackage{multirow}             
\usepackage{array}                
\usepackage{tikz}                 
\usetikzlibrary{quotes,angles,arrows,decorations.markings} 
\usepackage{makecell} 

\usepackage{lipsum}

\usepackage{xcolor}               
\usepackage{color}                
\usepackage{textcomp}             
\usepackage{units}                

\newcommand{\dd}{\mathrm{d}}
\newcommand{\e}{\mathrm{e}}
\newcommand{\ii}{\mathrm{i}}

\newcommand{\MA}{\mathrm{MA}}
\newcommand{\met}{\mathrm{met}}
\newcommand{\Kerr}{\mathrm{K}}
\newcommand{\order}{\mathcal{O}}
\newcommand{\be}{\begin{equation}}
\newcommand{\ee}{\end{equation}}
\newcommand{\Be}{\begin{eqnarray}}
\newcommand{\Ee}{\end{eqnarray}}

\def\e{\mbox{\rm e}}

\newcommand{\mincir}{\raise
-3.truept\hbox{\rlap{\hbox{$\sim$}}\raise4.truept\hbox{$<$}\ }}
\newcommand{\magcir}{\raise
-3.truept\hbox{\rlap{\hbox{$\sim$}}\raise4.truept\hbox{$>$}\ }}

\newcolumntype{Y}{>{\centering\arraybackslash}X}
\providecommand{\U}[1]

\usepackage[dvips]{epsfig}        
\usepackage[dvips]{graphicx}      
\usepackage{hyperref}             
\hypersetup{
    colorlinks=true,              
    breaklinks=true,              
    citecolor=blue,               
    linkcolor=[rgb]{0,0.5,0.9},   
    urlcolor=red,                 
    filecolor=green               
}

\newcommand{\ie}{\begin{equation}}
\newcommand{\fe}{\end{equation}}
\newcommand{\se}{\begin{eqnarray}}
\newcommand{\ff}{\end{eqnarray}}

\begin{document}
\raggedbottom

\title{{How does geometry shape quantum emission in rotating bumblebee gravity?}}


\author{A. A. Ara\'{u}jo Filho}
\email{dilto@fisica.ufc.br}
\affiliation{Departamento de Física, Universidade Federal da Paraíba, Caixa Postal 5008, 58051--970, João Pessoa, Paraíba,  Brazil.}
\affiliation{Departamento de Física, Universidade Federal de Campina Grande Caixa Postal 10071, 58429-900 Campina Grande, Paraíba, Brazil.}
\affiliation{Center for Theoretical Physics, Khazar University, 41 Mehseti Street, Baku, AZ-1096, Azerbaijan.}

\date{\today}

\begin{abstract}

{

We compare propagation and static emission estimates in a rotating metric geometry and a stationary \textit{metric--affine} bumblebee geometry. Their distinct constructions and spatial asymptotics are kept, and normalized Killing times define the physical frequencies and angular velocities. Direct invariant calculations identify the \textit{metric--affine} outer root and open exterior axes as curvature singularities for generic rotation. The obstruction at bifurcation surface poles concerns the smoothness of the bumblebee field independently. Its mixed meridional component couples angular channels; finite transverse energy fixes the adopted axial condition and defines a self adjoint stationary operator on bounded exterior annuli. For the metric geometry, we construct normalized radiative scalar modes with specified Unruh data, evaluate the horizon frequency transformation, and include spontaneous superradiant emission through the radial scattering map. We prove positivity of the axisymmetric potential and compute numerical transmission factors and a partial scalar energy spectrum alongside an analytical lower bound. In the static comparison at equal canonical energy, $T_{\met}=T_{\mathrm{Sch}}$ and $T_{\MA}=(1-X/4)T_{\mathrm{Sch}}$. The static emission estimates use the stated Stefan--Boltzmann model. 
}

\end{abstract}


\maketitle

\clearpage

\tableofcontents


\section{Introduction}
\label{sec:introduction}

Lorentz symmetry is one of the central principles underlying general relativity and the Standard Model of particle physics. Nevertheless, several candidates for a fundamental theory allow this symmetry to be broken at sufficiently high energies. In particular, spontaneous Lorentz symmetry breaking was first identified in string theory when tensor fields acquire nonvanishing vacuum expectation values \cite{KosteleckySamuel1989,KosteleckySamuel1989Gravity}. At low energies, possible departures from local Lorentz invariance can be described systematically within the Standard Model Extension \cite{ColladayKostelecky1998,Kostelecky2004,BaileyKostelecky2006}. Bumblebee gravity provides one of its simplest dynamical realizations: a vector field $B_\mu$ develops a nonzero vacuum expectation value $b_\mu$ through a potential $V(B_\mu B^\mu\mp b^2)$ and couples nonminimally to curvature. The theory preserves general covariance at the level of the action, while the vacuum selects a preferred spacetime direction \cite{BluhmKostelecky2005,BluhmFungKostelecky2008,BluhmEtAl2008,BertolamiParamos2005}.

Black holes have played a central role in the development of this model. After the early analysis of bumblebee vacuum configurations in Ref.~\cite{BertolamiParamos2005}, Casana \textit{et al.} obtained the first exact Schwarzschild--like black hole sourced by a radial spacelike vacuum expectation value \cite{Casana2018}. Its geometry differs from Schwarzschild mainly through a constant deformation of the radial sector, but this modification already affects the bending of light, the Shapiro delay, and the advance of perihelion, leading to strong Solar--System restrictions on the Lorentz--violating parameter. This solution was subsequently generalized in several directions. Static black holes with an effective cosmological constant were constructed \cite{MalufNeves2021} (and also wormholes \cite{Magalhaes:2025nql,Lessa:2025kln}); solutions containing a global monopole or a topological defect were found \cite{GulluOvgun2022}; {and black holes in higher dimensional asymptotically dS and AdS spacetimes were obtained \cite{DingChenFu2022,DingEtAl2023AdS}.} {Solutions with nonminimal couplings have also been proposed} \cite{Zhu:2026vae}. An Einstein--bumblebee--dilaton model obtained through dimensional reduction produced charged and thermodynamically stable configurations \cite{LessaSilva2023}, while an Einstein--Gauss--Bonnet extension supplied further exact black holes and modified phase structures \cite{DingChenFu2022}.

The static sector is considerably richer than the original one--parameter radial--vacuum geometry. Xu, Liang, and Shao found two families of spherical vacuum solutions, including configurations with a temporal bumblebee component and nonvanishing field strength, and studied their compatibility with Solar--System and black hole imaging data \cite{XuLiangShao2023}. {An analytical family with four parameters was also reported} in Ref.~\cite{Xu2023FourParameter}. Solutions away from the minimum of the bumblebee potential were investigated analytically and numerically in Ref.~\cite{BaileyMurrayWalterCardona2025}, revealing Schwarzschild--AdS and Reissner--Nordström limits, naked singularities, repulsive regions, and rapidly varying near-source fields. Purely timelike vacuum expectation values were considered in Ref.~\cite{LiZhu2025}, where nontrivial curved solutions were shown to require a special value of the vacuum norm. More recently, exact asymptotically flat black holes with a temporal bumblebee field were obtained in Ref.~\cite{YangMaiLiangShao2026}. Solutions with two independent components of $b_\mu$ and lightlike or spacelike vacuum expectation values were derived {in Refs.}~\cite{LiuWuWeiLiu2025VEV,Zhu:2025fiy}. In addition, a noncommutative extension of the static bumblebee black hole has also been obtained by implementing a Moyal deformation of the underlying geometry \cite{AraujoFilhoEtAl2026Noncommutative}.

Matter--supported configurations provide another important part of the current state of the art. A nonminimal interaction between the electromagnetic and bumblebee fields led to exact Reissner--Nordström--like and Reissner--Nordström--(A)dS--like black holes, together with their slowly rotating charged counterparts \cite{LiuGuoWeiLiu2025}. Exact dyonic solutions with electric and magnetic charges, general horizon topology, and Taub--NUT extensions were subsequently constructed in Refs.~\cite{ChenLiu2025TaubNUT,LiLiangMa2026}. Bumblebee extensions of the Kerr--Newman--Taub--NUT--(A)dS family can also be generated from closed constant-norm vacuum one--forms associated with geodesics of the seed geometry \cite{Ovcharenko2026}. The electromagnetic sector has been enlarged beyond Maxwell theory: charged static and slowly rotating solutions coupled to conformal ModMax electrodynamics were obtained in Ref.~\cite{SekhmaniEtAl2025ModMax}, while a general class of nonlinear--electrodynamic sources, including Born--Infeld--type models and marginally regular configurations, was considered in Ref.~\cite{LiLu2026NLED}. ModMax black holes surrounded by a cloud of strings and charged bumblebee black holes supported by string clouds have also been used to investigate horizon structure, thermodynamics, optical observables, greybody factors, and emission spectra \cite{AhmedKalaAlBadawi2026,BelchiorEtAl2026ModMax}.

These geometries have motivated a broad phenomenological program. For the Schwarzschild--like solution, the principal signatures include weak and finite--distance lensing, circular timelike orbits, perihelion precession, time delay, black hole thermodynamics, and accretion \cite{Casana2018,OvgunJusufiSakalli2019,LiOvgun2020,GomesMalufAlmeida2020,EslamPanah:2025zcm}. Scalar, electromagnetic, fermionic, and gravitational perturbations have been employed to calculate quasinormal frequencies, effective potentials, greybody factors, and stability domains \cite{OliveiraDantasAlmeida2021,GogoiGoswami2022,MalikEtAl2024,LiuEtAl2024Isospectrality}. {The dS, AdS, global monopole, and higher dimensional solutions} have been confronted with black hole shadows, EHT measurements, strong and weak deflection angles, thermodynamic phase transitions, and Hawking-radiation sparsity \cite{UniyalKanziSakalli2023,LinJiangZhai2023,PantigEtAl2025,SinghEtAl2025}. For charged backgrounds, photon and massive--particle motion, shadow morphology, strong lensing, quasinormal modes, coupled gravitoelectromagnetic perturbations, greybody transmission, periodic orbits, and gravitational waveforms have been investigated in Refs.~\cite{XuEtAl2025ChargedDynamics,SinghEtAl2025ChargedQNMs,LiEtAl2025ChargedQNMs,ShiEtAl2026Periodic}.

A conceptually different realization is obtained when the metric and affine connection are treated as independent variables \cite{DAmbrosio:2021zpm,Heisenberg:2018vsk,Heisenberg:2023lru} (and the equivalence principle could be violated \cite{Battista:2026rtl}). In \textit{metric--affine} bumblebee gravity, the independent connection can be solved algebraically in terms of an auxiliary metric, whereas the physical metric is related to it through a bumblebee--dependent disformal transformation \cite{DelhomEtAl2021,DelhomEtAl2022PLB,DelhomEtAl2022JCAP}. Radiative corrections and the one--loop effective action of this formulation have also been examined \cite{DelhomEtAl2022PLB,DelhomEtAl2022JCAP,LehumEtAl2024}. The first static vacuum black hole was derived in Ref.~\cite{AraujoFilho2023PRD}. Its phenomenology includes thin--disk accretion, weak and strong gravitational lensing, black hole shadows, neutrino propagation, quasinormal modes, greybody bounds, ringdown waveforms, scattering, Hawking sparsity, and evaporation \cite{LambiaseEtAl2023,HassanabadiEtAl2024,HeidariEtAl2024Scattering,Gao2024,JhaRahaman2024,JhaRahaman2025Constraints}. A direct comparison between particle creation and evaporation in the metric and \textit{metric--affine} static solutions was performed in Ref.~\cite{AraujoFilho2025JCAP}. The exact stationary and axisymmetric \textit{metric--affine} solution was obtained in Ref.~\cite{AraujoFilho2024JCAP}, and its geodesic structure, radial acceleration, thermodynamic quantities, phenomenological bounds, and shadow were subsequently explored \cite{AraujoFilho2024JCAP,NascimentoEtAl2026}.

Rotation is indispensable if bumblebee black holes are to be compared with astrophysical compact objects. It introduces inner and outer horizons, stationary limit surfaces, an ergoregion, frame dragging, rotational superradiance, and an angular chemical potential in the Hawking distribution. It also allows strong-field observables such as the innermost stable circular orbit, nodal and periastron precession frequencies, the deformation of the shadow, and relativistic disk spectra to probe sectors that are absent in spherical symmetry. Nevertheless, rotating bumblebee solutions have different theoretical statuses and should not be placed on equal footing. The Kerr--like metric of Ref.~\cite{Ding2020} was initially presented as an exact solution, but the bumblebee equation restricts its validity and supports, at most, an appropriate slow--rotation interpretation \cite{DingChen2021,MalufMuniz2022}. A consistent first--order rotating solution was obtained directly from the field equations in Ref.~\cite{DingChen2021}. Exact rotating gradient vacuum geometries were later generated through rank--one disformal deformations of Kerr in Ref.~\cite{PoulisSoares2022}. {Other solutions include} a Kerr--Sen--like construction \cite{JhaRahaman2021}, the exact rotating BTZ--like black hole \cite{DingEtAl2023BTZ} (and its features within quantum entanglement harvesting \cite{Liu:2025bpp}), slowly rotating charged and ModMax solutions \cite{LiuGuoWeiLiu2025,SekhmaniEtAl2025ModMax}, the exact \textit{metric--affine} geometry \cite{AraujoFilho2024JCAP}, and the five--dimensional equal angular momentum solutions with and without a cosmological constant \cite{ChenShenLiu2026}. {A stealth configuration in which} the metric remains exactly Kerr while the bumblebee field is nontrivial has also been identified \cite{XuMaiLiang2026StealthKerr}.

{A further rotating metric was constructed through a corrected Newman--Janis procedure~\cite{KumarIslamGhosh2026}. We examine its field equation status in Sec.~\ref{subsec:status}. The regularity results of~\cite{GuoFan2026} concern two independent obstructions. Horizon and axial symmetry force a smooth one form of nonzero constant norm to fail at a pole of a regular nondegenerate bifurcation surface. Independently, variation of the axial conicity obstructs a $C^2$ metric extension along an open axis. We test the field and metric conditions separately for the geometry studied here.}

The proposed rotating geometries have already been used to study several gravitational signatures. These include thin--disk fluxes and accretion efficiencies \cite{LiuDingJing2020}, finite--distance and strong gravitational lensing \cite{LiOvgun2020,KuangOvgun2022,MangutEtAl2023}, shadows in vacuum and plasma environments \cite{Ding2020,WangWei2022,IslamGhoshMaharaj2024}, scalar superradiant instability \cite{JiangLinZhai2021}, quasiperiodic oscillations \cite{WangChenJing2022}, X-ray reflection spectroscopy \cite{GuEtAl2022}, neutrino--antineutrino energy deposition \cite{KhodadiEtAl2023}, and scalar quasinormal modes \cite{LiuFangJingWang2023,ChenPanJing2023}.  At the same time, the strong degeneracy between the spin and Lorentz--violating parameters found in X-ray and shadow analyses shows that no single observable is sufficient for a model--independent constraint \cite{GuEtAl2022,IslamGhoshMaharaj2024}.

The quantum emission sector is less complete. For static bumblebee black holes, Hawking radiation has been studied through semiclassical tunneling, generalized uncertainty principle corrections, Bogoliubov coefficients, greybody bounds, energy spectra, sparsity, and evaporation lifetimes \cite{KanziSakalli2019,SakalliYoruk2023,GogoiGoswami2022,JhaRahaman2024,AraujoFilho2025JCAP}. Charged, cosmological, topological defect, and ModMax configurations have supplied additional examples in which Lorentz violation modifies the Hawking temperature, transmission probability, and energy emission rate \cite{SinghEtAl2025ChargedQNMs,BelchiorEtAl2026ModMax,AhmedKalaAlBadawi2026}. In rotating backgrounds, greybody radiation, scalar quasinormal modes, superradiant scattering, and approximate tunneling temperatures have been calculated for selected Kerr--like, Kerr--Sen--like, and slowly rotating geometries \cite{DingChen2021,JhaRahaman2021,KanziSakalli2021,JiangLinZhai2021}. Some of these results inherit the consistency limitations of the rotating metric employed, and the interpretation of the corresponding Hawking factor requires particular care in the superradiant interval, where the signed absorption probability must accompany the thermal occupation number \cite{Page1976,Page1976Rotating,MalufMuniz2022}.

{A comparison of scalar propagation and static emission estimates in bumblebee geometries requires explicit regularity, charge, and state assumptions.} Several questions remain open. The stationary Killing vector must first be normalized at spatial infinity before frequencies, angular velocities, and temperatures belonging to different geometries can be compared. One must then determine whether the horizon possesses a uniform surface gravity, whether the scalar wave equation separates into independent angular channels, how the off-diagonal meridional component $g_{r\theta}$ affects propagation, and whether a local tunneling factor can be promoted to a global Hawking spectrum. These questions cannot be settled by transferring the static expressions directly to the rotating metrics.

{We compare the corrected Newman--Janis geometry~\cite{KumarIslamGhosh2026} and the stationary \textit{metric--affine  } geometry~\cite{AraujoFilho2024JCAP}. Their physical metrics and bumblebee profiles are examined in Sec.~\ref{subsec:status}. The earlier Ding geometry~\cite{Ding2020} is excluded from the quantitative calculation.}
{The comparison keeps each geometry's vacuum profile and asymptotic structure explicit. We first establish the time and charge conventions and examine invariant regularity. We then study the scalar wave operator, normalized radiative modes, and numerical scattering, with the rotating metric--affine results restricted to the punctured exterior. Static emission and observational estimates retain the assumptions stated in Secs.~\ref{sec:evaporation} and \ref{sec:bounds}.}

{\section{The two rotating bumblebee geometries}
\label{sec:solutions}}

We use the signature $(-,+,+,+)$ and natural units $G=c=\hbar=k_{\mathrm B}=1$. The two Lorentz-violating parameters are not identified: $\ell$ belongs to the {metric geometry}, whereas $X$ controls the \textit{metric--affine} geometry. The azimuthal wave number is denoted by $m$, and the angular multipole number is denoted by $L$, avoiding confusion with $\ell$.


\subsection{{Rotating metric geometry}}
\label{subsec:metric-solution}

The rotating geometry used in Ref.~\cite{KumarIslamGhosh2026} can be written as
\begin{align}
\dd s_{\met}^{2} ={}&-\left(1-\frac{2\mathcal{M}(r)r}{\Sigma}\right)\dd t^{2} +\frac{\Sigma}{\Delta_{\ell}}\dd r^{2} +\Sigma\dd\theta^{2}  +\frac{\mathcal{A}_{\ell}\sin^{2}\theta}{\Sigma}\dd\phi^{2} -\frac{4a\mathcal{M}(r)r\sin^{2}\theta}{\Sigma}\dd t\,\dd\phi , \label{eq:metric-line-element}\\ \Sigma={}&r^{2}+a^{2}\cos^{2}\theta,\qquad \mathcal{A}_{\ell}=(r^{2}+a^{2})^{2} -a^{2}\Delta_{\ell}\sin^{2}\theta , \nonumber\\ \mathcal{M}(r)={}& \frac{M\left(1+\frac{r\ell}{2M}\right)}{1+\ell}, \qquad \Delta_{\ell} =a^{2}+\frac{r(r-2M)}{1+\ell}.
\nonumber
\end{align}
{With $k=1+\ell>0$, the asymptotic value $g_{tt}\to-1/k$ gives the normalized time} 
\begin{equation} 
\tau_{\met}=\frac{t}{\sqrt{k}}, \qquad \partial_{\tau_{\met}}=\sqrt{k}\,\partial_t. \label{eq:metric-time-normalization} \end{equation} 
{Frequencies, angular velocities, temperatures, fluxes, and static evolution estimates refer to the corresponding normalized asymptotic time.}

The static limit of Eq.~\eqref{eq:metric-line-element}, after Eq.~\eqref{eq:metric-time-normalization}, is naturally the Schwarzschild--like bumblebee metric
\begin{equation}
\dd s_{\met}^{2}\big|_{a=0} =-f(r)\dd\tau_{\met}^{2} +\frac{k}{f(r)}\dd r^{2}+r^{2}\dd\Omega^{2}, \qquad f(r)=1-\frac{2M}{r}. 
\label{eq:metric-static-limit} 
\end{equation}


\subsection{Rotating solution in the \textit{metric--affine} formulation}
\label{subsec:ma-solution}

The \textit{metric--affine} solution was obtained by solving the field equations in an auxiliary Einstein frame and mapping the Kerr geometry to the physical metric \cite{AraujoFilho2024JCAP}. {Define}
\begin{equation}
\alpha=1+\frac{3X}{4},\qquad \beta=1-\frac{X}{4},\qquad q=\sqrt{\alpha\beta},
\nonumber
\end{equation}
with $\alpha>0$ and $\beta>0$, or $-\frac{4}{3}<X<4$. The physical line element is
\begin{align}
\dd s_{\MA}^{2} ={}&-\frac{\Delta-a^{2}\sin^{2}\theta}{q\Sigma}\dd t^{2} -\frac{4aMr\sin^{2}\theta}{q\Sigma}\dd t\,\dd\phi +\frac{(r^{2}+a^{2})^{2}-a^{2}\Delta\sin^{2}\theta}        {q\Sigma}\sin^{2}\theta\,\dd\phi^{2} \nonumber\\ &+\frac{a^{2}\cos^{2}\theta+r^{2}\alpha/\beta}         {q\Delta}\dd r^{2} +\frac{r^{2}+a^{2}\cos^{2}\theta\,\alpha/\beta}         {q}\dd\theta^{2} +\frac{2rXa\cos\theta} {\sqrt{\alpha}\,\beta^{3/2}\sqrt{\Delta}}\dd r\,\dd\theta , \label{eq:ma-line-element}\\ \Delta={}&r^{2}-2Mr+a^{2},\qquad \Sigma=r^{2}+a^{2}\cos^{2}\theta .
\nonumber
\end{align}

{The mixed component $g_{r\theta}=rXa\cos\theta/(\sqrt\alpha\,\beta^{3/2}\sqrt\Delta)$ agrees with Eqs.~\eqref{eq:ma-inverse-thth}--\eqref{eq:ma-det} and couples radial and polar scalar propagation.}

{The asymptotic value $g_{tt}\to-1/q$ gives}
\begin{equation}
\tau_{\MA}=\frac{t}{\sqrt q}, \qquad \partial_{\tau_{\MA}}=\sqrt q\,\partial_t.
\label{eq:ma-time-normalization}
\end{equation}
{For later use, define}
\begin{align}
\mathcal{R}(r,\theta) &=\beta r^{2}+\alpha a^{2}\cos^{2}\theta, \label{eq:Rdef}\\ \mathcal{P}(r,\theta) &=\alpha r^{2}+\beta a^{2}\cos^{2}\theta.
\label{eq:Pdef}
\end{align}
Direct inversion of the $(r,\theta)$ block gives
\begin{align}
g^{rr} =\frac{\beta\mathcal{R}\Delta}{q\Sigma^{2}}, \, \, \, g^{r\theta} =-\frac{rXa\cos\theta}{\Sigma^{2}} \sqrt{\frac{\beta}{\alpha}}\sqrt{\Delta}, \, \, \,  g^{\theta\theta} =\frac{\beta\mathcal{P}}{q\Sigma^{2}}.
\label{eq:ma-inverse-thth}
\end{align}
The determinant identities
\begin{align}
g_{rr}g_{\theta\theta}-g_{r\theta}^{2}  =\frac{\Sigma^{2}}{\beta^{2}\Delta}, \, \, \, \det(g_{\mu\nu}) =-\frac{\Sigma^{2}\sin^{2}\theta}{\alpha\beta^{3}}, \, \, \,  \sqrt{-g}=\frac{\Sigma\sin\theta} {\sqrt{\alpha}\,\beta^{3/2}},
\label{eq:ma-det}
\end{align}
provide useful algebraic checks.

The relation between the physical and auxiliary metrics can be written as
\begin{equation}
g_{\mu\nu} =\frac{1}{q}h_{\mu\nu} +\frac{\xi}{\alpha}B_\mu B_\nu, \qquad \Gamma^\lambda{}_{\mu\nu} =\left\{{}^{\,\lambda}_{\mu\nu}\right\}_{h}.
\label{eq:disformal-map}
\end{equation}
Using the convention $Q_{\lambda\mu\nu}\equiv\nabla^{(\Gamma)}_\lambda g_{\mu\nu}$, the \textit{nonmetricity} tensor is 
\begin{equation}
Q_{\lambda\mu\nu} =\frac{\xi}{\alpha} \left[ \left(\nabla^{(h)}_\lambda B_\mu\right)B_\nu +B_\mu\left(\nabla^{(h)}_\lambda B_\nu\right) \right].
\label{eq:nonmetricity-tensor}
\end{equation}
In other words, $X$ labels the disformal deformation, while the local \textit{nonmetricity} also depends on derivatives of the bumblebee profile. This distinction is important when deciding whether a quantum observable contains a direct $Q_{\lambda\mu\nu}$ coupling or only an $X$--dependent metric effect.


{

\subsection{Status and regularity of the geometries}
\label{subsec:status}
}

The two metrics do not have the same theoretical status. The \textit{metric--affine} geometry in Eq.~\eqref{eq:ma-line-element} was obtained from the corresponding field equations \cite{AraujoFilho2024JCAP}. The geometry in Eq.~\eqref{eq:metric-line-element} was generated from the static metric by a modified Newman--Janis prescription \cite{KumarIslamGhosh2026}. Such a construction does not, by itself, prove that the rotating metric and a rotating bumblebee vacuum satisfy the complete metric field equations and the bumblebee equation. {We use it as a prescribed physical metric.}

{

In the standard vacuum sector $B_{\mu\nu}=0$ and $V'=0$, the bumblebee equation gives $R^\mu{}_{\nu}B^\nu=0$. For $k=1+\ell$ and $M(r)=(M+\ell r/2)/k$, the mixed Ricci tensor has the two double eigenvalues~\cite{TorresFayos2017}
\[
\lambda_1 = \frac{\ell a^2\cos^2\theta}{k\Sigma^2}, \qquad \lambda_2 = \frac{\ell r^2}{k\Sigma^2}, \qquad \det(R^\mu{}_{\nu}) = \frac{\ell^4a^4r^4\cos^4\theta}{k^4\Sigma^8}.
\]
For $a \ell \neq 0$, this tensor is invertible on open exterior sets away from the equatorial plane, forcing $B^\mu = 0$ there. Notice that Eq.~(\ref{eq:metric-line-element}) lies outside the nonzero constant norm sector with vanishing field strength. Its rotating results describe the prescribed metric. Configurations with $B_{\mu\nu}\neq0$ or $V'\neq0$ involve different field equations and are outside this check. }

{{Our comparison concerns two different geometries.} At fixed metric parameter $M$, the choice $\ell=3X/4$ matches the leading static temperature corrections. Their distinct constructions and spatial asymptotics remain part of the comparison. In other words, the calculated differences characterize these particular metric families.}

{

The regularity of Eq.~\eqref{eq:ma-line-element} can be examined directly through its disformal representation,
\begin{equation}
g_{\mu\nu} = \frac{1}{q}\left(h_{\mu\nu}+\frac{X}{\beta}\omega_\mu\omega_\nu\right), \qquad \omega = \frac{r}{\sqrt{\Delta}}\,\mathrm{d}r+a\cos\theta\,\mathrm{d}\theta, 
\label{eq:ma_disformal_regularity} 
\end{equation}
where $h_{\mu\nu}$ is the auxiliary Kerr metric and $h^{\mu\nu}\omega_\mu\omega_\nu=1$. The underlying bumblebee profile is
\begin{equation} 
B_\mu \mathrm{d}x^\mu = \widetilde b\,\omega, \qquad \widetilde b = b\,\alpha^{1/4}\beta^{-3/4}, \qquad g^{\mu\nu}B_\mu B_\nu=b^2>0. 
\label{eq:ma_bumblebee_regularity} 
\end{equation}
For a nonextremal configuration, $M>|a|$, direct evaluation of the physical metric curvature gives
\begin{equation} \left.R[g]\right|_{\theta=\pi/2} = -\frac{aX\,\Delta'(r_+)}{r_+^3\sqrt{\Delta}}\sqrt{\frac{\beta}{\alpha}}+O(1), \qquad r\rightarrow r_+^+.
\label{eq:ma_ricci_divergence} 
\end{equation}
Since $\Delta'(r_+)=r_+-r_->0$, the outer root is a curvature singular boundary when $aX\neq0$. This divergence occurs at a nonpolar latitude and excludes a smooth extension across the complete surface.

In the connection gauge of Eq.~(\ref{eq:disformal-map}), $\Gamma^\lambda{}_{\mu\nu} = \{^\lambda{}_{\mu\nu}\}_h$, the affine Ricci tensor vanishes and the auxiliary Kerr curvature remains finite at $r=r_+$. The bumblebee norm and its vanishing field strength are also finite. A derivative invariant supplies a separate test. Defining $H_{\mu\nu}=\nabla^{(h)}_\mu\omega_\nu$ and
\begin{equation} 
Q_\lambda = g^{\mu\nu}Q_{\lambda\mu\nu}, \qquad \mathring Q_{\lambda\mu\nu} = Q_{\lambda\mu\nu}-\frac14 Q_\lambda g_{\mu\nu}, 
\end{equation}
{For $\Sigma_+=r_+^2+a^2\cos^2\theta$, we obtain $Q_\lambda=0$ and}
\begin{equation} 
\mathring Q_{\lambda\mu\nu}\mathring Q^{\lambda\mu\nu} = \frac{2X^2}{q}H_{\mu\nu}H^{\mu\nu}_{(h)} = \frac{X^2r_+^2(r_+-r_-)}{2q\Sigma_+^2(r-r_+)}+O\!\left((r-r_+)^{-1/2}\right). 
\label{eq:ma_nonmetricity_divergence} 
\end{equation}
Indices on $\mathring Q_{\lambda\mu\nu}$ are raised with the physical metric. Its trace free definition removes the pure trace contribution of a projective transformation. Equation~\eqref{eq:ma_nonmetricity_divergence} identifies a further obstruction to smoothness of the complete configuration for $X\neq0$.

The profile in Eq.~\eqref{eq:ma_bumblebee_regularity} inherits stationarity and axial symmetry and has strictly positive constant norm. A proposed smooth nondegenerate bifurcate extension satisfies the hypotheses of the fixed point argument in Ref.~\cite{GuoFan2026}. At a bifurcation surface pole, horizon and axial invariance force a smooth one form to vanish, contradicting $B_\mu B^\mu=b^2>0$. This argument is independent of the affine field equations. Its scope concerns nondegenerate bifurcate horizons; future only and extremal horizons lie outside that particular theorem. For the present nonextremal rotating geometry, Eq.~\eqref{eq:ma_ricci_divergence} already establishes the metric obstruction. Moreover, $B_\theta=\widetilde b\,a\cos\theta$ fails the axial smoothness condition $B_\theta=O(\sin\theta)$ when $a\neq0$.

The open exterior axes possess a separate metric curvature obstruction, independent of the bumblebee smoothness argument at bifurcation surface poles. Writing $F=g_{\phi\phi}$, their invariant conicity for an azimuthal period $2\pi$ is
\begin{equation} 
c^2(r) = \lim_{\theta\rightarrow0,\pi}\frac{g^{\mu\nu}\partial_\mu F\,\partial_\nu F}{4F} = 1-\frac{Xa^2}{\alpha(r^2+a^2)}, \qquad \frac{\mathrm{d}c^2}{\mathrm{d}r} = \frac{2Xa^2r}{\alpha(r^2+a^2)^2}.
\label{eq:ma_axial_conicity} 
\end{equation}
This calculation includes the complete meridional inverse metric. On open axial segments with $r>r_+$, the bounded azimuthal quotient data permit application of the varying conicity criterion of Ref.~\cite{GuoFan2026}. With $\rho$ denoting transverse proper distance and $z$ proper distance along the axis,
\begin{equation} R_{\hat\rho\hat\phi\hat z\hat\phi} = -\frac{\partial_z\ln c}{\rho}+O(1). \label{eq:ma_axial_curvature} \end{equation}
Writing $D=r^2+a^2$, direct evaluation of the meridional Schur complement on either axis gives
\begin{equation} H_{\rm ax}(r) = \left.\left(g_{\theta\theta}-\frac{g_{r\theta}^2}{g_{rr}}\right)\right|_{\theta=0,\pi} = \frac{qD^2}{\beta(\alpha r^2+\beta a^2)}, \end{equation}
which reproduces the conicity in Eq.~(\ref{eq:ma_axial_conicity}). The complete Levi--Civita Kretschmann scalar of the physical metric satisfies
\begin{equation} {\cal K}[g] = \frac{C_{\rm ax}(r)}{\sin^2\theta}+O(\sin^{-1}\theta), \qquad \theta\rightarrow0,\pi, \end{equation}
where
\begin{equation} C_{\rm ax}(r) = \frac{4\beta X^2a^2}{\alpha}\left[\frac{\Delta}{D^4}+\frac{M^2r^2(r^2-a^2)^2}{D^6\Delta}\right]. \label{eq:complete_axial_coefficient} \end{equation}
For $\alpha,\beta>0$, $r>r_+$, and $aX\neq0$, the first term is strictly positive and the second is nonnegative. In this case, $C_{\rm ax}(r)>0$ throughout the open exterior. Both axes are scalar curvature singularities, independently of the azimuthal period. This axial divergence vanishes for $X=0$ or $a=0$. The limit is taken at fixed $r>r_+$; notice that the nonpolar outer root singularity and the obstruction at bifurcation surface poles are established separately above.

The rotating \textit{metric--affine} results describe local WKB propagation on $r>r_+$, $0<\theta<\pi$, where the fields are smooth on compact subsets. The radial expansions above assume nonextremality, while the exterior axial obstruction persists in the rotating extremal limit. For $a=0$, the physical metric admits a regular metric horizon, although Eq.~\eqref{eq:ma_nonmetricity_divergence} still diverges for $X\neq0$. The static thermal calculation refers to canonically coupled test fields on the extended physical metric.}

{

\subsection{Asymptotic charges and physical parameters}
\label{subsec:charges}

We use the normalized asymptotic times and fix $\phi\sim\phi+2\pi$ in the charges, areas, and angular modes. The leading static spatial geometries are
\[
\mathrm{d}\sigma_{\rm met}^2 \simeq \mathrm{d}\varrho_{\rm met}^2+\frac{\varrho_{\rm met}^2}{k}\,\mathrm{d}\Omega^2, \qquad \mathrm{d}\sigma_{\rm MA}^2 \simeq \mathrm{d}\varrho_{\rm MA}^2+\frac{\beta}{\alpha}\,\varrho_{\rm MA}^2\mathrm{d}\Omega^2,
\]
\[
\varrho_{\rm met}=\sqrt{k}\,r, \qquad \varrho_{\rm MA}=\sqrt{\frac{\alpha}{q\beta}}\,r.
\]
Their asymptotic solid angles are $4\pi/k$ and $4\pi\beta/\alpha$. We define the charges from the gravitational action, with fixed coupling parameters and zero energy assigned to $M=0$.

For the \textit{metric--affine} configuration, the gravitational momentum density satisfies
\[
\mathcal P^{\mu\nu}=\sqrt{-g}\left(\beta g^{\mu\nu}+\xi B^\mu B^\nu\right)=\sqrt{-h}\,h^{\mu\nu}.
\]
On $\Gamma=\{h\}$, the gravitational symplectic potential and Noether charge form coincide with their Einstein expressions for the auxiliary Kerr metric. The vector contribution vanishes along the permitted variations because $B_{\mu\nu}=0$. The Hamiltonian variations are~\cite{IyerWald1994}
\[
\delta H_\xi=\int_{S_\infty}\left(\delta\boldsymbol Q_\xi-\xi\mathbin{\cdot}\boldsymbol\Theta\right).
\]
With $\partial_{\tau_{\rm MA}}=\sqrt q\,\partial_t$, this gives
\begin{equation}
\delta E_{\rm MA}=\sqrt q\,\delta M, \qquad \delta J_{\rm MA}=\delta(Ma), \qquad E_{\rm MA}=\sqrt q\,M, \qquad J_{\rm MA}=Ma.
\end{equation}
The charge forms have the smooth auxiliary Kerr limits at the poles, and their excised polar contributions vanish. The physical metric keeps the axial curvature obstruction established in Sec.~II~C. The fixed azimuthal period defines the angular integrals throughout the calculation.

The canonical dimensionless spin and outer root condition are
\begin{equation}
j_{\rm MA} = \frac{J_{\rm MA}}{E_{\rm MA}^2}=\frac{a}{qM}, \qquad q^2j_{\rm MA}^2\leq1.
\end{equation}
Writing $E=E_{\rm MA}$, $J=J_{\rm MA}$, $j=J/E^2$, $s=\sqrt{1-q^2j^2}$, and $u=1+s$, the change of variables is
\[M = \frac{E}{\sqrt q}, \qquad a=\frac{\sqrt q\,J}{E}, \qquad r_+=\frac{Eu}{\sqrt q}, \qquad E^2\geq q|J|.\]
The normalized angular velocity and exterior WKB scale become
\[\Omega_{\rm MA}(E,j)=\frac{q^2j}{2Eu}, \qquad T_{\rm MA}(E,j,\theta)=\frac{qs}{4\pi Eu}\sqrt{\frac{\beta u^2+\alpha q^2j^2\cos^2\theta}{\alpha\left(u^2+q^2j^2\cos^2\theta\right)}}.\]
The limiting area is
\[A_{\rm MA}(E,j) = \frac{8\pi E^2u}{q^2}\int_0^1\sqrt{\frac{1+(\alpha/\beta)\zeta x^2}{1+\zeta x^2}}\,\mathrm{d}x, \qquad \zeta=\frac{q^2j^2}{u^2}.\]
These quantities refer to the outer root geometry and its exterior limits, with the regularity status specified in Sec.~II~C.

For comparison with Kerr at the same $E$ and $J$, let $s_K=\sqrt{1-j^2}$. In this manner, we have
\[\Omega_K=\frac{j}{2E(1+s_K)}, \qquad T_K=\frac{s_K}{4\pi E(1+s_K)}, \qquad A_K=8\pi E^2(1+s_K), \qquad |j|\leq1.\]
The common charge domain is $|j|\leq\min(1,q^{-1})$. The nonzero $\ell$ rotating metric in Eq.~(\ref{eq:metric-line-element}) remains a prescribed geometry under the vacuum check above; its curves are presented in terms of $(M,a)$ and $\chi=a/M$. The static comparison uses the canonical energies of both specified static configurations.
Also, Figs.~\ref{hawking}--\ref{b2}, \ref{greybody}, \ref{f3}, and \ref{f4} compare identical metric parameters $(M,a)$; $\chi=a/M$ denotes their parameter ratio.

}


\section{Horizons, stationary limit surfaces, and angular velocities}
\label{sec:horizons}

\subsection{{Metric geometry}}

The roots of $\Delta_{\ell}=0$ are
$r_{\pm}^{(\met)} =M\pm d_{\ell}$, and $  d_{\ell}=\sqrt{M^{2}-ka^{2}}$. The black hole domain is $M^{2}\geq ka^{2}$, and extremality occurs for $M^{2}=ka^{2}$. Positive $\ell$ decreases the outer horizon at fixed $(M,a)$ and lowers the largest permitted spin.

The stationary limit surfaces follow from $g_{\tau_{\met}\tau_{\met}}=0$:
\begin{equation}
r_{\mathrm{sl},\pm}^{(\met)}(\theta)
=M\pm\sqrt{M^{2}-ka^{2}\cos^{2}\theta}.
\label{eq:metric-stationary-limit}
\end{equation}
At the poles, $r_{\mathrm{sl},+}^{(\met)}=r_{+}^{(\met)}$, while at the equator $r_{\mathrm{sl},+}^{(\met)}=2M$. The coordinate thickness of the ergoregion is therefore modified by $\ell$.

The local frame--dragging angular velocity in the original $t$ coordinate is
\begin{equation}
\omega_t(r,\theta) =-\frac{g_{t\phi}}{g_{\phi\phi}} =\frac{a\left(r^{2}+a^{2}-\Delta_{\ell}\right)} {(r^{2}+a^{2})^{2}-a^{2}\Delta_{\ell}\sin^{2}\theta}.
\label{eq:metric-frame-dragging}
\end{equation}
At the outer horizon,
\begin{equation}
\Omega_{\met} =\sqrt{k}\,\omega_t(r_+,\theta) =\frac{\sqrt{k}\,a} {\left(r_{+}^{(\met)}\right)^{2}+a^{2}}.
\label{eq:metric-horizon-angular-velocity}
\end{equation}
The factor $\sqrt{k}$ is required by the normalized time \eqref{eq:metric-time-normalization}. Notice that, at least in the metric formulation, at fixed $(M,a)$, a positive $\ell$ shifts the outer horizon and the stationary limit surface inward, lowers the extremal spin, and increases the horizon angular velocity defined with respect to the normalized asymptotic time.


{\subsection{Metric--affine geometry}}

{For the \textit{metric--affine} geometry, the roots of $\Delta=0$ are} 
$r_{\pm}^{(\MA)} =M\pm d$ , $d=\sqrt{M^{2}-a^{2}}$. {Their coordinate locations coincide with the Kerr roots.} The stationary limit surfaces are also unchanged: \begin{equation} 
r_{\mathrm{sl},\pm}^{(\MA)}(\theta) =M\pm\sqrt{M^{2}-a^{2}\cos^{2}\theta}. \label{eq:ma-stationary-limit} 
\end{equation}

\begin{figure}
    \centering
    \includegraphics[scale=0.6]{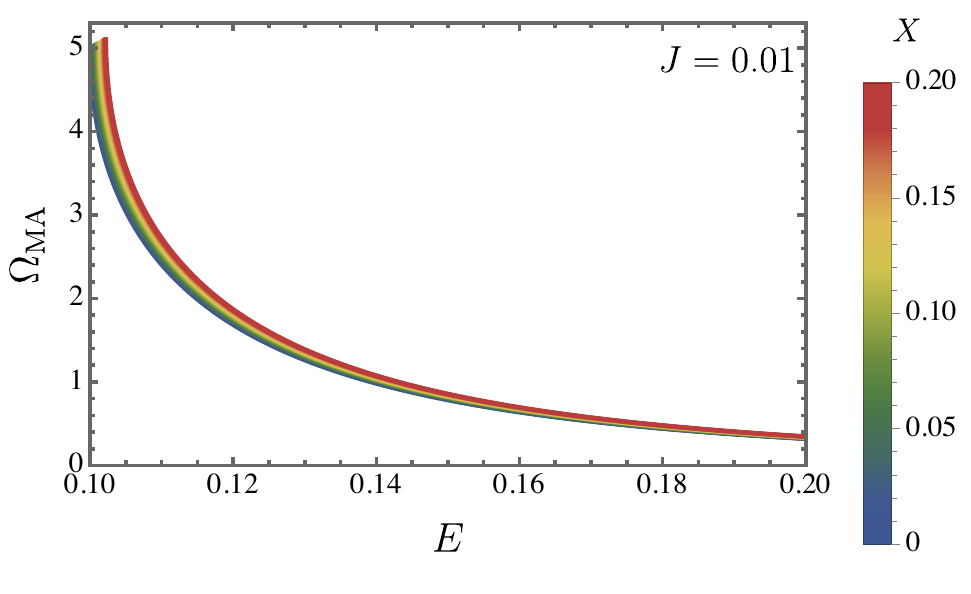}
    \caption{{Normalized angular velocity of the \textit{metric--affine} outer root as a function of canonical energy $E$, at fixed canonical angular momentum $J$. The $X=0$ curve is the Kerr result at the same charges. Each curve is restricted to $E^2\geq q|J|$.}}
    \label{angularvelovity}
\end{figure}

The coordinate horizon angular velocity is $a/(r_+^2+a^2)$. With the normalization in Eq.~\eqref{eq:ma-time-normalization}, it becomes \begin{equation} 
\Omega_{\MA} =\frac{\sqrt q\,a} {\left(r_{+}^{(\MA)}\right)^{2}+a^{2}}. \label{eq:ma-horizon-angular-velocity} \end{equation} 
{Figure~\ref{angularvelovity} compares the \textit{metric--affine} angular velocity at fixed canonical angular momentum $J$ with the Kerr result at the same charges. The curves use the canonical energy $E$ and the domain $E^2\geq q|J|$. The exterior interpretation of the \textit{metric--affine} outer root is specified in Sec.~II~C.}

{\section{Surface gravity, temperature scales, and area geometry}
\label{sec:thermodynamics}}

\subsection{Normalized surface gravity of the {metric geometry}}

Let $\chi=\partial_t+\omega_{H,t}\partial_\phi$ be the horizon generator in the unnormalized coordinate. {Direct evaluation of}
\begin{equation}
\kappa_t^{2} =\lim_{r\rightarrow r_+} \frac{g^{\mu\nu}\partial_\mu(-\chi^2) \partial_\nu(-\chi^2)} {4(-\chi^2)}
\label{eq:kappa-limit}
\end{equation}
{gives}
\begin{equation}
\kappa_{\met,t} =\frac{\Delta_{\ell}'(r_+)} {2\left[\left(r_+^{(\met)}\right)^2+a^2\right]} =\frac{d_{\ell}} {k\left[\left(r_+^{(\met)}\right)^2+a^2\right]}.
\label{eq:metric-kappa-coordinate}
\end{equation}
The derivative $\Delta_{\ell}'(r)=2(r-M)/k$ is essential. Omitting $1/k$ would treat $\Delta_{\ell}$ as the Kerr radial function.

Rescaling the horizon generator according to Eq.~\eqref{eq:metric-time-normalization} yields
\begin{align}
\kappa_{\met} &=\sqrt{k}\,\kappa_{\met,t} =\frac{d_{\ell}} {\sqrt{k}\left[\left(r_+^{(\met)}\right)^2+a^2\right]}, \label{eq:metric-kappa-physical}\\ T_{\met} &=\frac{\kappa_{\met}}{2\pi} =\frac{d_{\ell}} {2\pi\sqrt{k}\left[\left(r_+^{(\met)}\right)^2+a^2\right]}.
\label{eq:metric-temperature}
\end{align}
The static result is
\begin{equation}
T_{\met}(a=0)=\frac{1}{8\pi M\sqrt{k}}, \label{eq:metric-static-temperature} \end{equation} as obtained directly from Eq.~\eqref{eq:metric-static-limit}. Positive $\ell$ lowers the temperature and moves the extremal boundary to smaller $a/M$.


{\subsection{Surface gravity scale of the metric--affine geometry}
\label{subsec:ma-kappa}}

{For the stationary Killing vector}
\begin{equation}
\chi_t=\partial_t+\frac{a}{r_+^2+a^2}\partial_\phi,
\label{eq:ma-coordinate-generator}
\end{equation}
{the $(t,\phi)$ block of the metric is $1/q$ times the Kerr block. In the exterior limit $r\downarrow r_+$,}
\begin{equation}
-\chi_t^2 =\frac{\Sigma_+\Delta'(r_+)} {q(r_+^2+a^2)^2}(r-r_+) +\order\!\left((r-r_+)^2\right),
\label{eq:ma-chi-expansion}
\end{equation}
where
\begin{equation}
\Sigma_+=r_+^2+a^2\cos^2\theta,\qquad \mathcal{R}_+=\beta r_+^2+\alpha a^2\cos^2\theta.
\label{eq:ma-horizon-functions}
\end{equation}
{The limiting scales are}
\begin{align}
\kappa_{\MA,t}(\theta) &=\frac{\Delta'(r_+)}{2(r_+^2+a^2)} \sqrt{\frac{\mathcal{R}_+}{\alpha\Sigma_+}}, \label{eq:ma-kappa-coordinate}\\ \kappa_{\MA}(\theta) &=\sqrt q\,\kappa_{\MA,t}(\theta) =\frac{d}{r_+^2+a^2} \sqrt{\frac{q\mathcal{R}_+}{\alpha\Sigma_+}}, \label{eq:ma-kappa-physical}\\ T_{\MA}(\theta) &=\frac{d}{2\pi(r_+^2+a^2)} \sqrt{\frac{q\mathcal{R}_+}{\alpha\Sigma_+}}.
\label{eq:ma-temperature}
\end{align}
{In an equivalent manner,}
\begin{equation}
T_{\MA}(\theta) =T_{\Kerr} \sqrt{\frac{q\left(\beta r_+^2+\alpha a^2\cos^2\theta\right)} {\alpha\left(r_+^2+a^2\cos^2\theta\right)}}, \qquad T_{\Kerr}=\frac{d}{2\pi(r_+^2+a^2)}.
\label{eq:ma-temperature-compact}
\end{equation}

To first order in $X$,
\begin{equation}
\frac{T_{\MA}(\theta)}{T_{\Kerr}} =1+\frac{X\left(a^2\cos^2\theta-3r_+^2\right)} {8\left(r_+^2+a^2\cos^2\theta\right)} +\order(X^2).
\label{eq:ma-temperature-expansion}
\end{equation}
The static limit is independent of $\theta$:
\begin{equation}
T_{\MA}(a=0) =\frac{1}{8\pi M} \frac{\beta^{3/4}}{\alpha^{1/4}}.
\label{eq:ma-static-temperature}
\end{equation}
{For small $X$,}
\begin{equation}
\frac{T_{\MA}(a=0)}{T_{\mathrm{Sch}}}
=1-\frac{3X}{8}+\order(X^2).
\label{eq:ma-static-temperature-expansion}
\end{equation}

{

For the standard radial static vacuum, the canonical energy is $E_{\rm met}=\sqrt{k}\,M$, with $k=1+\ell$~\cite{An2024}. Comparing the two static geometries at equal canonical energy $E$ gives
\[
T_{\rm met}(E) = \frac{1}{8\pi E}, \qquad T_{\rm MA}(E) = \frac{\beta}{8\pi E}, \qquad A_{\rm met}(E) = \frac{16\pi E^2}{k}, \qquad A_{\rm MA}(E) = \frac{16\pi E^2}{\alpha\beta}.
\]
{The illustrative choice $\ell=3X/4$ is discussed in Appendix~\ref{sec:time-normalization}.} At fixed canonical energy, the metric temperature is independent of $\ell$, while $T_{\rm MA}/T_{\rm met}=1-X/4$.
}

\begin{figure}
    \centering
    \includegraphics[scale=0.6]{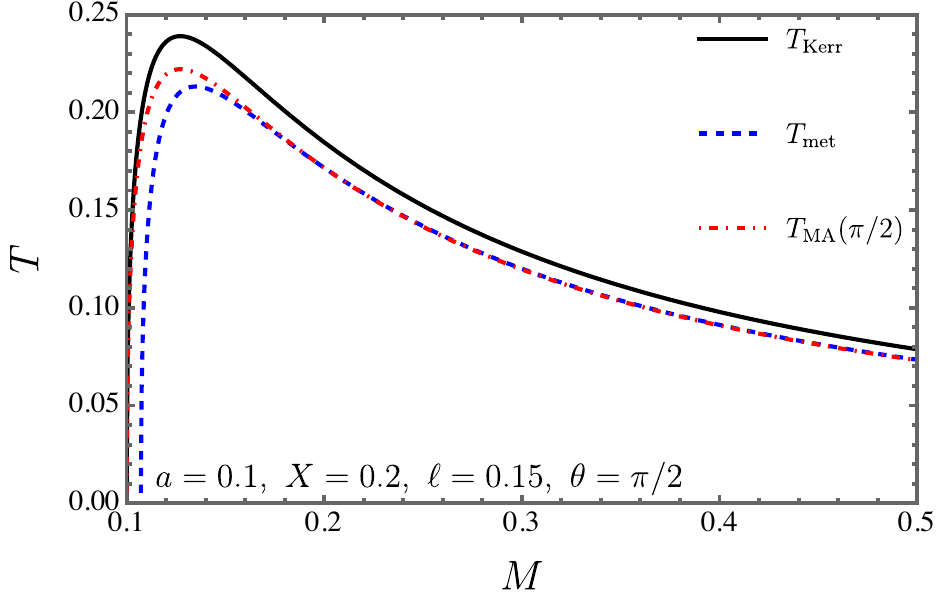}
    \caption{Comparison of the Kerr temperature $T_{\mathrm{Kerr}}$, the metric Hawking temperature $T_{\mathrm{met}}$, and the {equatorial \textit{metric--affine} exterior WKB scale} $T_{\MA}(\pi/2)$ as functions of $M$, for $a=0.1$, $X=0.2$, and $\ell=3X/4=0.15$. Lorentz violation suppresses the temperature relative to Kerr.}
    \label{hawking}
\end{figure}

{For $aX\neq0$, the exterior WKB scale in Eq.~\eqref{eq:ma-temperature} varies with latitude. Its polar to equatorial ratio is}
\begin{equation}
\frac{T_{\MA}(0)}{T_{\MA}(\pi/2)}
=\sqrt{1+\frac{Xa^2}{\beta(r_+^2+a^2)}}.
\label{eq:temperature-anisotropy}
\end{equation}
{For $aX\neq0$, Eqs.~(\ref{eq:ma-kappa-coordinate})--(\ref{eq:temperature-anisotropy}) define exterior limiting scales, with $T_{\rm MA}(\theta)$ specifying the radial WKB exponent. In addition, notice that a smooth nondegenerate Euclidean completion would require a single period $\beta_E$ satisfying $\beta_E\kappa=2\pi$ over the connected horizon. The invariant divergences in Eqs.~(\ref{eq:ma_ricci_divergence}) and (\ref{eq:ma_nonmetricity_divergence}) exclude the necessary regular extension, independently of the chosen period. The regular bifurcate horizon setting of the zeroth law is absent.} In Fig. \ref{hawking}, we compare the Kerr temperature, the Hawking temperature of the {metric geometry}, and the {equatorial exterior WKB scale of the \textit{metric--affine} geometry} as functions of $M$. We adopt $a=0.1$, $X=0.2$, and {the choice} $\ell=3X/4=0.15$. All curves vanish at their respective extremal boundaries, attain a maximum, and subsequently decrease as $M$ increases. Lorentz violation suppresses the temperature relative to Kerr. Moreover, the {metric geometry} exhibits the shifted extremal boundary $M_{\mathrm{ext}}=\sqrt{1+\ell}\,a$, whereas the \textit{metric--affine} geometry retains the Kerr value $M_{\mathrm{ext}}=a$.  

{For the illustrated positive deformations, the metric temperature and equatorial \textit{metric--affine} WKB scale exchange their ordering between the metric extremal boundary and the limit $M/a\to\infty$.}

{

\subsection{Cross section areas}
\label{subsec:areas}

We use $\mathscr A=A/4$ as a geometrical normalization. For the metric geometry,
\begin{equation} A_{\met}=4\pi\left[\left(r_+^{(\met)}\right)^2+a^2\right], \qquad \mathscr A_{\met}=\frac{A_{\met}}4. \label{eq:metric-area} \end{equation}
For the \textit{metric--affine} geometry, define the exterior limiting area by
\begin{equation} A_{\MA}=\lim_{\epsilon\downarrow0}\lim_{r_0\downarrow r_+}\int_{\epsilon}^{\pi-\epsilon}\!\dd\theta\int_0^{2\pi}\!\dd\phi\,\sqrt{g_{\theta\theta}g_{\phi\phi}}\big|_{t=\mathrm{constant},\,r=r_0}. \label{eq:ma-area-limit} \end{equation}
For $\alpha,\beta>0$ and $r_+>0$, the area density has an integrable polar limit, giving
\begin{equation} A_{\MA}=\frac{4\pi(r_+^2+a^2)}q\int_0^1\sqrt{\frac{\beta r_+^2+\alpha a^2u^2}{\beta(r_+^2+a^2u^2)}}\,\dd u, \qquad \mathscr A_{\MA}=\frac{A_{\MA}}4. \label{eq:ma-area-exact} \end{equation}
Its small $X$ expansion and static limit are
\begin{equation} A_{\MA}=4\pi(r_+^2+a^2)\left[1+X\left(\frac14-\frac{r_+}{2a}\tan^{-1}\frac a{r_+}\right)\right]+\order(X^2). \label{eq:ma-area-expansion} \end{equation}
\begin{equation} A_{\MA}(a=0)=\frac{16\pi M^2}q. \label{eq:ma-static-area} \end{equation}
For $aX\neq0$, the curvature singularities established in Sec.~II~C prevent the complete limiting surface, including its poles, from defining a regular Noether charge surface. Its finite improper area remains a geometrical quantity. The entropy of the full theory follows from the complete nonminimally coupled Lagrangian~\cite{Wald1993,IyerWald1994}; here we study area derivatives, with no equilibrium ensemble or gravitational first law assigned to them. Figure~\ref{entropy} compares $\mathscr A$ at identical metric parameters.
\begin{figure}
\centering
\includegraphics[scale=0.6]{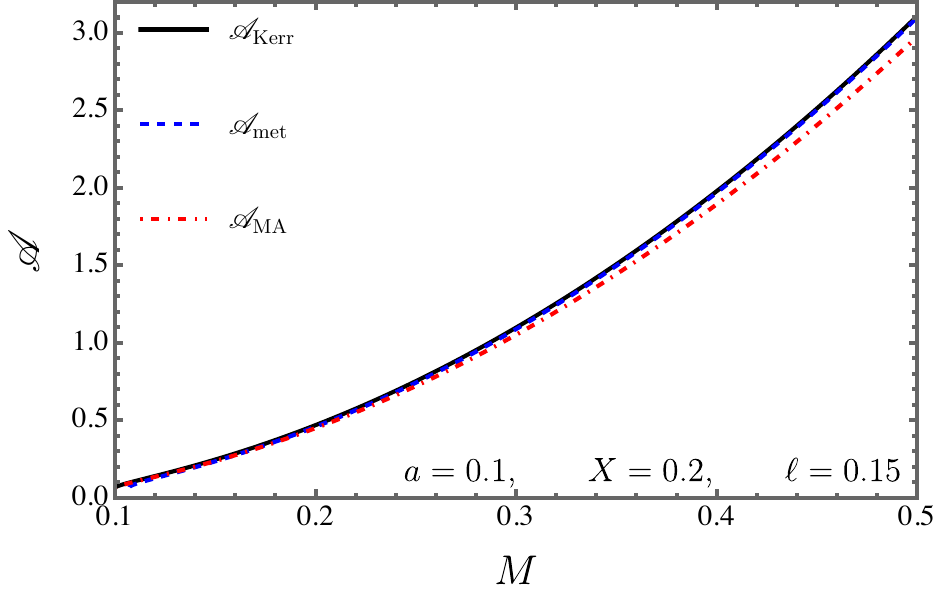}
\caption{{Geometrically normalized areas $\mathscr A=A/4$ for Kerr, the metric geometry, and the \textit{metric--affine} exterior limit, as functions of the metric parameter $M$, at $a=0.1$, $X=0.2$, and $\ell=3X/4=0.15$.}}
\label{entropy}
\end{figure}

\subsection{Geometrical area derivatives}

At fixed $a$ and coupling parameters, we define a derivative along each {radius parametrized} curve by
\begin{equation} \mathcal D^{(A)}=\frac{\dd\mathscr A/\dd r}{\dd\ln T/\dd r}. \label{eq:area-scale-derivative} \end{equation}
For the metric geometry, $r=r_+^{(\met)}$ and $\Delta_\ell(r)=0$ give
\begin{equation} M=\frac{r^2+ka^2}{2r}, \qquad T_{\met}(r)=\frac{r^2-ka^2}{4\pi\sqrt{k}\,r(r^2+a^2)}, \qquad \mathscr A_{\met}(r)=\pi(r^2+a^2). \label{eq:metric-M-r} \end{equation}
The derivatives are
\begin{equation} \frac{\dd\mathscr A_{\met}}{\dd r}=2\pi r, \qquad \frac{\dd T_{\met}}{\dd r}=\frac{-r^4+(1+3k)a^2r^2+ka^4}{4\pi\sqrt{k}\,r^2(r^2+a^2)^2}. \label{eq:metric-temperature-derivative} \end{equation}
In this regard,
\begin{equation} \mathcal D^{(A)}_{\met,a}=\frac{2\pi r^2(r^2+a^2)(r^2-ka^2)}{-r^4+(1+3k)a^2r^2+ka^4}. \label{eq:metric-area-derivative} \end{equation}
For $a\neq0$, the positive stationary point of $T_{\met}(r)$ is
\begin{equation} \frac{r_{\mathrm{turn}}^2}{a^2}=\frac{1+3k+\sqrt{(1+3k)^2+4k}}2. \label{eq:metric-turning-radius} \end{equation}

For the \textit{metric--affine} exterior limit, let $r=r_+^{(\MA)}$, $\Sigma_\theta=r^2+a^2\cos^2\theta$, and $R_\theta=\beta r^2+\alpha a^2\cos^2\theta$. The radial WKB scale is
\begin{equation} T_{\MA}(r,\theta)=\frac{r^2-a^2}{4\pi r(r^2+a^2)}\sqrt{\frac{qR_\theta}{\alpha\Sigma_\theta}}. \label{eq:ma-temperature-r} \end{equation}
At fixed $a,X,\theta$, we write
\begin{equation} \mathcal I(r)=\int_0^1\sqrt{\frac{\beta r^2+\alpha a^2u^2}{\beta(r^2+a^2u^2)}}\,\dd u, \qquad \mathscr A_{\MA}(r)=\frac{\pi(r^2+a^2)}q\,\mathcal I(r). \label{eq:ma-area-integral} \end{equation}
\begin{equation} \mathcal I'(r)=\int_0^1\sqrt{\frac{\beta r^2+\alpha a^2u^2}{\beta(r^2+a^2u^2)}}\left[\frac{\beta r}{\beta r^2+\alpha a^2u^2}-\frac r{r^2+a^2u^2}\right]\dd u. \label{eq:ma-area-integral-derivative} 
\end{equation}
\begin{equation} 
\mathcal G_\theta(r)=\partial_r\ln T_{\MA}(r,\theta)=\frac{2r}{r^2-a^2}-\frac1r-\frac{2r}{r^2+a^2}+\frac{\beta r}{R_\theta}-\frac r{\Sigma_\theta}. 
\label{eq:ma-temperature-log-derivative} 
\end{equation}
The geometrical derivative is
\begin{equation} \mathcal D^{(A)}_{\MA,a}(\theta)=\frac\pi q\,\left(\frac{2r\mathcal I(r)+(r^2+a^2)\mathcal I'(r)}{\mathcal G_\theta(r)} \right). 
\label{eq:ma-local-response} 
\end{equation}
Its poles locate stationary points of the WKB scale along the specified fixed latitude curves. In the static limit,
\begin{equation} \mathcal D^{(A)}_{\met,a=0}=-8\pi M^2, \qquad \mathcal D^{(A)}_{\MA,a=0}=-\frac{8\pi M^2}q. \label{eq:static-area-derivatives} \end{equation}
These expressions and Fig.~\ref{heat} describe geometrical parameter derivatives; their signs and poles characterize the specified curves.

\begin{figure}
\centering
\includegraphics[scale=0.6]{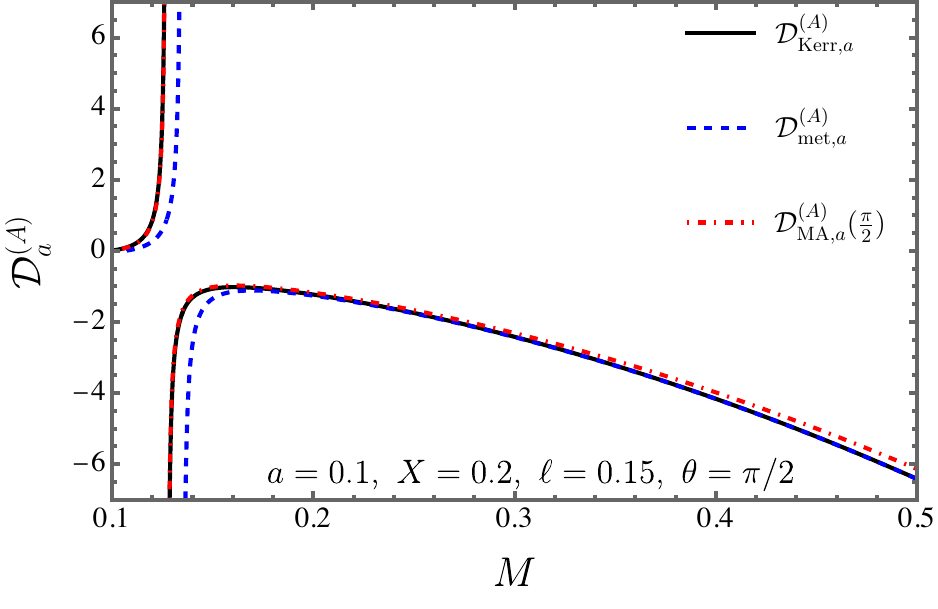}
\caption{{Geometrical derivatives $\mathcal D^{(A)}_{K,a}$, $\mathcal D^{(A)}_{\met,a}$, and $\mathcal D^{(A)}_{\MA,a}(\pi/2)$ along fixed $a$ curves, as functions of the metric parameter $M$, at $a=0.1$, $X=0.2$, and $\ell=3X/4=0.15$. The Kerr curve follows from $k=1$ in the metric expression.}}
\label{heat}
\end{figure}
}

{\section{Hamilton--Jacobi tunneling and occupation factors}
\label{sec:tunneling}}


\subsection{{Metric geometry}}

{

Let us consider $r_+=r_+^{(\met)}$ and consider a minimally coupled scalar of mass $\mu_{\mathrm p}$. The WKB form $\Phi=\mathcal A\exp(\ii I/\hbar)$ gives the leading Hamilton--Jacobi equation~\cite{Vanzo2011,aa2023implications}
\begin{equation} g^{\mu\nu}\partial_\mu I\,\partial_\nu I+\mu_{\mathrm p}^{\,2}=0. \label{eq:metric-HJ} \end{equation}
With the frequency conjugate to the normalized clock, write
\begin{equation} I=-\bar\omega t+m\phi+W(r)+\Theta(\theta), \qquad \bar\omega=\frac{\omega}{\sqrt k}. \label{eq:metric-action} \end{equation}
Separation gives $\mathcal K=[\Theta'(\theta)]^2+(m/\sin\theta-a\bar\omega\sin\theta)^2+\mu_{\mathrm p}^{\,2}a^2\cos^2\theta$ and the radial momenta
\begin{equation} W_\pm'(r)=\pm\frac{\sqrt{[(r^2+a^2)\bar\omega-am]^2-\Delta_\ell(\mu_{\mathrm p}^{\,2}r^2+\mathcal K)}}{\Delta_\ell}. \label{eq:metric-radial-momentum} \end{equation}
At a nonextremal outer horizon, $\Delta_\ell=\Delta_\ell'(r_+)(r-r_+)+O((r-r_+)^2)$. Analytic continuation retains the sign of $(r_+^2+a^2)\bar\omega-am$, distinguishing the superradiant and nonsuperradiant sectors. The prescription $r-r_+\to r-r_+-\ii0$ gives
\begin{equation} \operatorname{Im}W_\pm=\pm\pi\frac{(r_+^2+a^2)\bar\omega-am}{\Delta_\ell'(r_+)}. \label{eq:metric-imaginary-action} \end{equation}
The mass and angular separation terms leave the residue unchanged. Including the temporal contribution in the invariant tunneling prescription~\cite{ParikhWilczek2000,Angheben2005,Vanzo2011} gives
\begin{equation} \mathcal B_{\met}\equiv\frac{\mathcal P_{\mathrm{em}}}{\mathcal P_{\mathrm{abs}}}=\exp\!\left[-\frac{4\pi[(r_+^2+a^2)\bar\omega-am]}{\Delta_\ell'(r_+)}\right]=\exp\!\left[-\frac{\omega-m\Omega_{\met}}{T_{\met}}\right]. \label{eq:metric-tunneling-rate} \end{equation}
Here, $(r_+^2+a^2)\bar\omega-am=(r_+^2+a^2)(\omega-m\Omega_{\met})/\sqrt k$. The residue vanishes continuously at the superradiant threshold.
}

{

On the other hand, the tunneling factor fixes local detailed balance. Its exponent leaves the quantum state and correlations between channels unspecified. Let $\mathsf c=(i,j,L,m,p)$ label the species, frequency bin, angular numbers, and polarization, and define
\begin{equation} \eta_i=(-1)^{2s_i}, \qquad x_{\mathsf c}^{(\met)}=\frac{\omega_j-m\Omega_{\met}}{T_{\met}}. \label{eq:metric-mode-variables} \end{equation}
Under a thermal occupation ansatz, the Bose or Fermi factor for one mode is
\begin{equation} \left\langle n_{\mathsf c}\right\rangle_{\met}^{(H)}=\frac{1}{\exp[x_{\mathsf c}^{(\met)}]-\eta_i}. \label{eq:metric-occupation} \end{equation}
For bosons, this occupation ansatz is defined for $x_{\mathsf c}^{(\met)}>0$.
}

{For the scalar state specified in Sec.~\ref{subsec:quantum-state}, the outgoing occupation includes the signed absorption probability $\mathcal A_{\mathsf c}^{(\met)}(\omega)$. The Fermi expression retains the thermal assumption. The resulting Page form is} \cite{Page1976,Page1976Rotating}
{
\begin{equation} \left\langle n_{\mathsf{c}}\right\rangle_{\met}^{\mathrm{out}} = \frac{ \mathcal{A}_{\mathsf{c}}^{(\met)}(\omega) }{ \exp\left[ (\omega-m\Omega_{\met})/T_{\met} \right]-\eta_i }. \label{eq:metric-page-occupation} \end{equation}
}
For a nonsuperradiant mode, $\mathcal{A}_{\mathsf{c}}^{(\met)}$ is the ordinary transmission probability. For a bosonic superradiant mode, $\omega<m\Omega_{\met}$, both the denominator and the signed absorption probability are negative. Their ratio remains positive. The thermal factor alone must therefore not be interpreted as either a particle number or an emission probability.

The zero temperature limit also remains finite when the signed absorption probability is retained:
{
\begin{equation} \lim_{T_{\met}\rightarrow0^+} \left\langle n_{\mathsf{c}}\right\rangle_{\met}^{\mathrm{out}} = -\eta_i\, \mathcal{A}_{\mathsf{c}}^{(\met)}(\omega) \Theta\left(m\Omega_{\met}-\omega\right). \label{eq:metric-extremal-emission} \end{equation}
}
For bosons, this expression describes spontaneous emission in the superradiant interval. The corresponding continuum sums for the particle, energy, and angular momentum fluxes are given in Sec.~VII.


{\subsection{Metric--affine geometry}}

Let $r_+=r_+^{(\MA)}$. For a mode
\begin{equation}
I=-\bar\omega t+m\phi+W(r,\theta), \qquad \bar\omega=\frac{\omega}{\sqrt q},
\label{eq:ma-coordinate-frequency}
\end{equation}
{the leading Hamilton--Jacobi equation near the outer root gives}
\begin{equation}
\partial_rW_\pm \simeq \pm \sqrt{\frac{\alpha\Sigma_+}{\mathcal{R}_+}}\, \frac{(r_+^2+a^2)\bar\omega-am} {\Delta'(r_+)(r-r_+)}.
\label{eq:ma-Wprime}
\end{equation}
{With $\delta=(r-r_+)/r_+$, Eq.~\eqref{eq:ma-Wprime} gives $W_\pm=A_\pm(\theta)\ln\delta+O(1)$, where $A_\pm$ is the radial residue. In this regard, $\partial_\theta W_\pm=A_\pm^{\prime}(\theta)\ln\delta+O(1)$ at a generic latitude. Since $g^{r\theta}=\order(\sqrt{\Delta})$ and $\partial_rW=\order(\Delta^{-1})$, the mixed term is $\order(\Delta^{-1/2}\ln\delta)$, while the polar quadratic term is $\order((\ln\delta)^2)$. Both are subleading to the $\order(\Delta^{-1})$ terms that fix the residue.}

At a fixed nonpolar latitude,
\begin{equation}
\operatorname{Im}W_\pm(\theta) = \pm\pi \sqrt{\frac{\alpha\Sigma_+}{\mathcal{R}_+}}\, \frac{(r_+^2+a^2)\bar\omega-am} {\Delta'(r_+)}.
\label{eq:ma-imaginary-action}
\end{equation}
{
The formal radial contour factor is
\begin{equation} {\cal B}_{\rm MA}(\theta) \equiv \exp\!\left[-\frac{4\pi}{\Delta'(r_+)}\sqrt{\frac{\alpha\Sigma_+}{{\cal R}_+}}\left((r_+^2+a^2)\bar\omega-am\right)\right] = \exp\!\left[-\frac{\omega-m\Omega_{\rm MA}}{T_{\rm MA}(\theta)}\right]. \end{equation}
This factor specifies the logarithmic phase structure of the exterior Hamilton--Jacobi solution. Physical particle production depends on a specified quantum state and boundary conditions, as we should naturally expect.}

\begin{figure}
    \centering
    \includegraphics[scale=0.55]{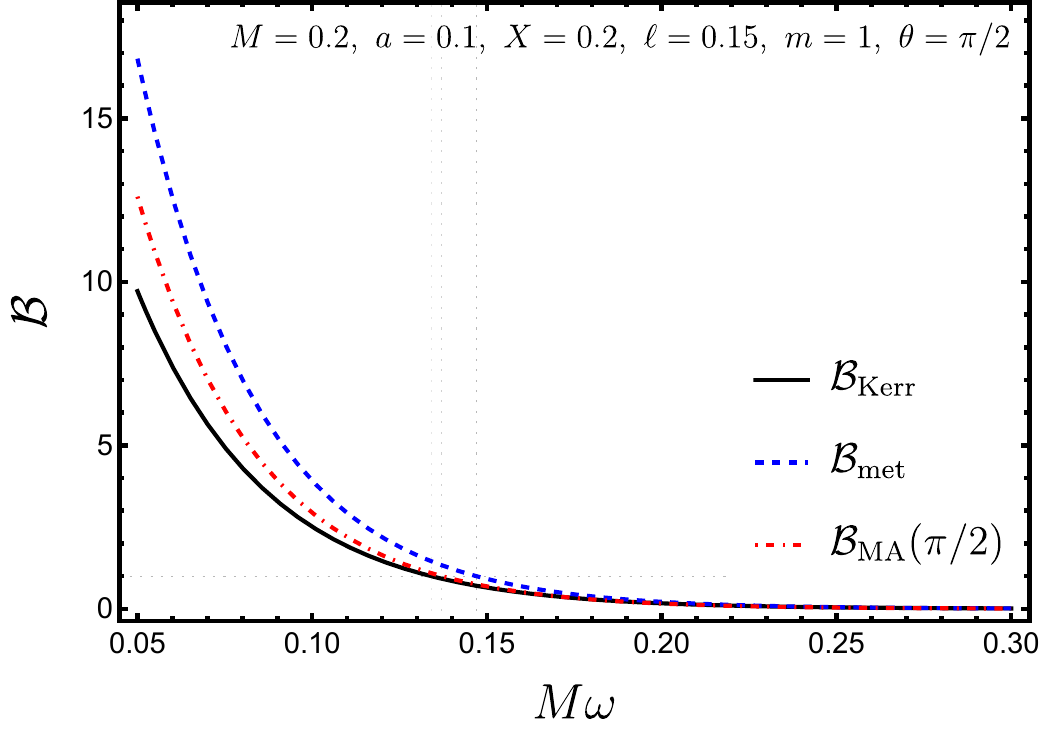}
    \caption{{Radial contour factors $\mathcal B_{\mathrm{Kerr}}$, $\mathcal B_{\mathrm{met}}$, and $\mathcal B_{\mathrm{MA}}(\pi/2)$ as functions of $M\omega$, for $M=0.2$, $a=0.1$, $m=1$, $X=0.2$, and $\ell=3X/4=0.15$. The dotted lines mark $\mathcal B=1$ and $M\omega=mM\Omega_H$. Values $\mathcal B>1$ identify negative corotating frequency.}}
    \label{b1}
\end{figure}

{Figure~\ref{b1} compares the radial contour factors of the Kerr, metric, and \textit{metric--affine} geometries as functions of $M\omega$.} We fix $M=0.2$, $a=0.1$, $m=1$, and adopt $X=0.2$ together with {the choice} $\ell=3X/4=0.15$. {The horizontal dotted line at $\mathcal B=1$ marks zero corotating frequency, while the vertical dotted lines indicate $M\omega=mM\Omega_H$.} {For these parameters, the thresholds are} $M\Omega_{\mathrm{Kerr}}\simeq0.1340$, $M\Omega_{\mathrm{MA}}\simeq0.1370$, and $M\Omega_{\mathrm{met}}\simeq0.1469$. Lorentz violation changes both the {normalized angular velocity and the temperature or exterior WKB scale}, producing distinct exponential slopes and shifting the onset of the nonsuperradiant regime. In particular, the {metric geometry} exhibits the largest displacement of the superradiant threshold. {Values $\mathcal B>1$ identify negative $\omega-m\Omega_H$; an amplification coefficient follows from the scattering problem.}

\begin{figure}
    \centering
    \includegraphics[scale=0.55]{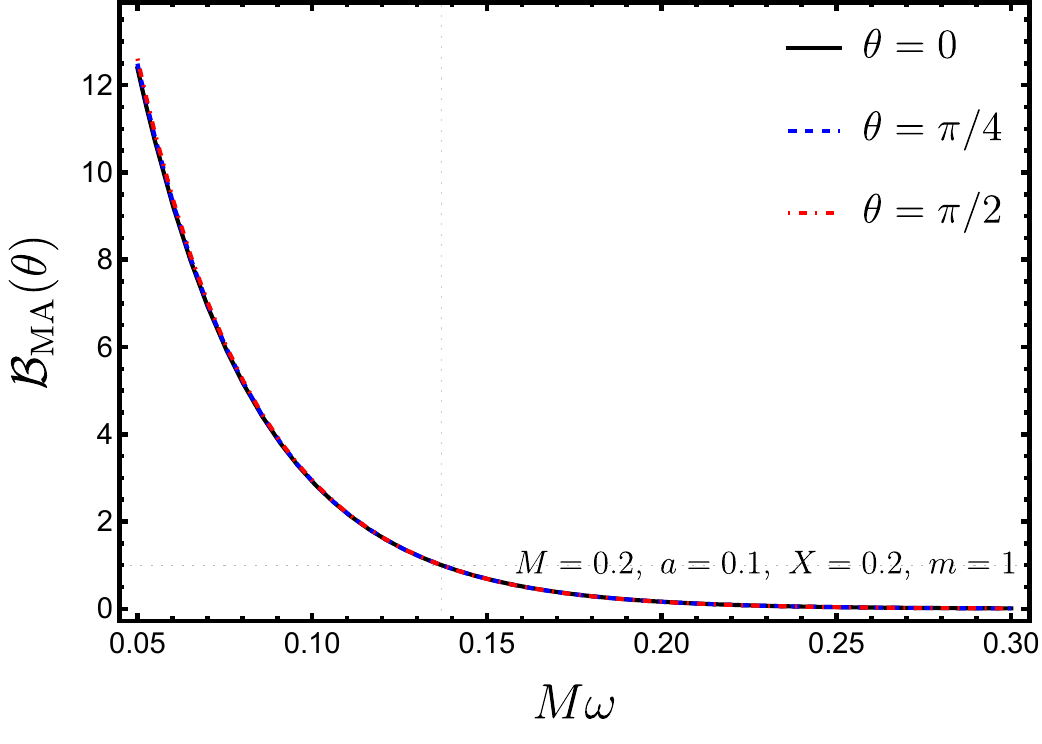}
    \caption{{\textit{Metric--affine} contour factor $\mathcal B_{\mathrm{MA}}(\theta)$ as a function of $M\omega$ at $\theta=\pi/4$ and $\pi/2$, together with the polar limit $\theta\downarrow0$, for $M=0.2$, $a=0.1$, $m=1$, and $X=0.2$. The curves meet at $M\omega=mM\Omega_{\mathrm{MA}}$. }}
    \label{b2}
\end{figure}

{Figure~\ref{b2} displays the latitude dependence of the \textit{metric--affine} contour factor at $\theta = \pi/4$ and $\pi/2$, together with its polar limit $\theta\downarrow0$.} Since the normalized angular velocity $\Omega_{\mathrm{MA}}$ is independent of $\theta$, the three curves meet at the common superradiant threshold $M\omega=mM\Omega_{\mathrm{MA}}\simeq0.1370$, where $\mathcal{B}_{\mathrm{MA}}=1$. For positive $X$, {the exterior WKB scales satisfy} $T_{\mathrm{MA}}(0)>T_{\mathrm{MA}}(\pi/4)>T_{\mathrm{MA}}(\pi/2)$. {Within the superradiant interval, the ordering is} $\mathcal{B}_{\mathrm{MA}}(\pi/2)>\mathcal{B}_{\mathrm{MA}}(\pi/4)>\mathcal{B}_{\mathrm{MA}}(0)$, whereas this ordering is reversed above the threshold. 

\begin{figure}
    \centering
    \includegraphics[scale=0.55]{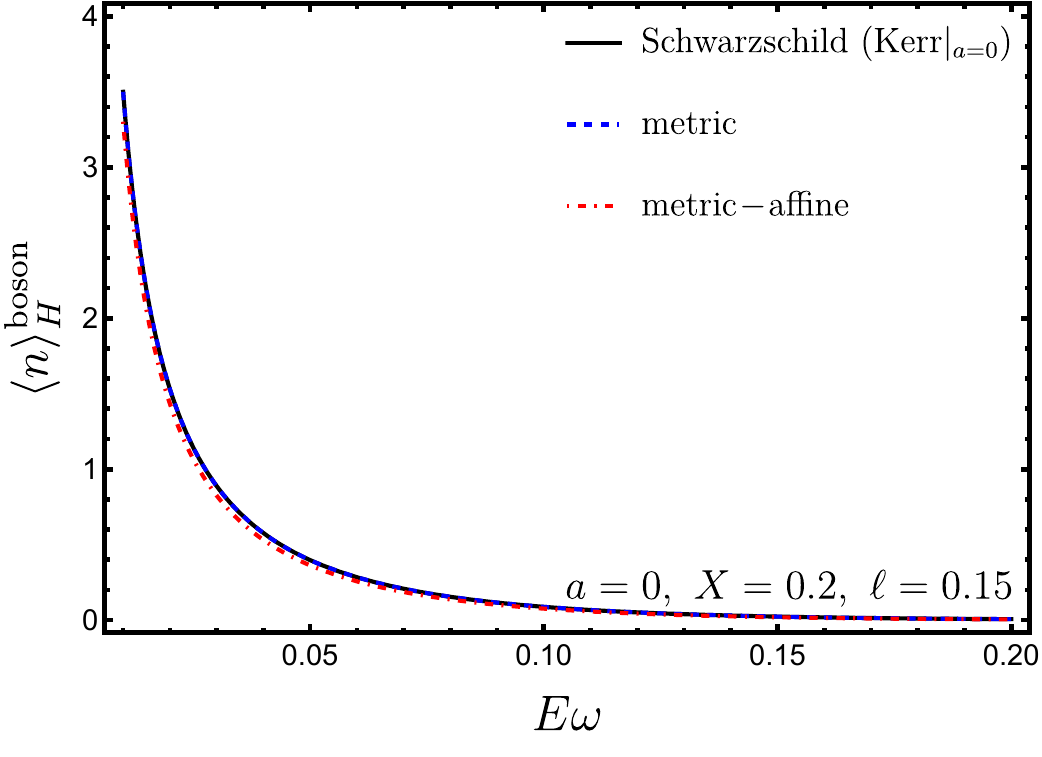}
    \caption{{Static Bose factors $[\exp(\omega/T)-1]^{-1}$ as functions of $E\omega$, at equal canonical energy, for $X=0.2$ and $\ell=3X/4$. The Schwarzschild and metric curves coincide. The curves illustrate an occupation ansatz with greybody factors omitted.}}
    \label{b3}
\end{figure}

{Figure~\ref{b3} compares the static Bose factors $[\exp(\omega/T)-1]^{-1}$ at equal canonical energy. The Schwarzschild and metric curves coincide, while the \textit{metric--affine} curve is lower for $X>0$. These are assumed occupation factors with greybody transmission omitted.}

{For $aX\neq0$, $\mathcal B_{\MA}(\theta)$ is an exterior contour factor at a fixed nonpolar latitude. The coordinate $\theta$ labels position and supplies no independent quantum channel. It is important to mention that a global occupation distribution does not follow from this factor.}

{

The limits $X=0$ and $a=0$ restore a uniform scale and separable angular modes. In the static limit,
\begin{equation} T_{\MA}^{(0)}=\frac{1}{8\pi M}\frac{\beta^{3/4}}{\alpha^{1/4}}, \qquad \mathcal B_{\MA}^{(a=0)}=\exp\!\left[-8\pi M\frac{\alpha^{1/4}}{\beta^{3/4}}\,\omega\right]. \label{eq:ma-static-tunneling} \end{equation}
{With the stated Unruh prescription, the outgoing static occupation is}
\begin{equation} \left\langle n_{\mathsf c}\right\rangle_{\MA,a=0}^{\mathrm{out}}=\frac{\mathcal A_{\mathsf c}^{(\MA,0)}(\omega)}{\exp[\omega/T_{\MA}^{(0)}]-\eta_i}. \label{eq:ma-static-page-occupation} \end{equation}
This conditional formula refers to test fields on the extended static physical metric.
}


\subsection{Direct and indirect \textit{nonmetricity} at WKB order}
\label{subsec:direct-indirect}

For a scalar field coupled canonically to the physical metric, the matter action may be written as
\begin{equation}
S_{\Phi} = -\frac{1}{2} \int\dd^4x\,\sqrt{-g} \left( g^{\mu\nu}\partial_\mu\Phi\partial_\nu\Phi +\mu_\Phi^2\Phi^2 \right).
\label{eq:canonical-scalar-action}
\end{equation}
Its field equation is
\begin{equation}
\left(\Box_g-\mu_\Phi^2\right)\Phi=0, \qquad \Box_g\Phi = \frac{1}{\sqrt{-g}} \partial_\mu \left( \sqrt{-g}\,g^{\mu\nu}\partial_\nu\Phi \right).
\label{eq:scalar-operator}
\end{equation}
Under this matter coupling prescription, the independent connection and $Q_{\lambda\mu\nu}$ do not appear explicitly. With $\Phi=A\exp(\ii I/\hbar)$, the leading WKB equation is
\begin{equation}
g^{\mu\nu}\partial_\mu I\partial_\nu I+\mu_\Phi^2=0.
\label{eq:scalar-eikonal}
\end{equation}
{For this matter coupling, the scalar WKB exponent and propagation depend on the physical metric $g_{\mu\nu}(X)$. The resulting differences characterize the chosen geometries. The axial curvature singularity is likewise a property of the physical metric, independently of any explicit matter coupling to $Q_{\lambda\mu\nu}$.}

For a spinor coupled to the independent connection, the Dirac equation may be written schematically as
\begin{equation}
\left( \ii\hbar\gamma^\mu D_\mu^{(\Gamma)} -\mu_\psi \right)\psi=0.
\label{eq:dirac-ma}
\end{equation}
The explicit factor of $\hbar$ is required to organize the WKB expansion. The independent connection, including its nonmetric part, enters the spinor transport equation directly \cite{DelhomEtAl2022PLB,DelhomEtAl2022JCAP}. Nevertheless, substituting $\psi=u\exp(\ii I/\hbar)$ gives, at leading order,
\begin{equation}
g^{\mu\nu}\partial_\mu I\partial_\nu I+\mu_\psi^2=0.
\label{eq:dirac-eikonal}
\end{equation}
The connection dependent part of the standard spinor covariant derivative contributes at the next WKB order. {{The leading fermionic exponent contains the same exterior WKB scale $T_{\MA}(\theta)$ as the scalar calculation.}}


{

\subsection{Scalar modes, state choice, and quantum correlations}
\label{subsec:quantum-state}

We now construct the scalar frequency transformation independently of the radial contour. We consider the canonically coupled real massless scalar of Sec.~VI, with $\phi\sim\phi+2\pi$, and a nonextremal metric geometry. The incoming state is specified by vacuum data at past infinity and positive affine frequency data on the extended past horizon, following the Unruh prescription~\cite{Unruh1974,Unruh1976}. The fermionic occupation formula above keeps its stated thermal assumption.


\subsubsection{Normalized radiative modes of the metric geometry}

Write $R=r^2+a^2$, $R_+=r_+^2+a^2$, $\tau=\tau_{\met}$, $\Omega_H=\Omega_{\met}$, and $\kappa=\kappa_{\met}>0$. The coordinate change
\begin{equation} \dd v=\dd t+\frac{R}{\Delta_\ell}\dd r, \qquad \dd\psi=\dd\phi+\frac{a}{\Delta_\ell}\dd r \label{eq:state-ingoing-chart} \end{equation}
gives $g_{vr}=1$, $g_{r\psi}=-a\sin^2\theta$, and $g_{rr}=0$. The remaining coefficients are finite at $r=r_+$, and $\det g=-\Sigma^2\sin^2\theta$. Then, the metric has a regular extension across its outer horizon; the usual polar coordinates cover its smooth axis with overlapping angular charts. The retarded coordinate change, with the two radial signs reversed, covers the past horizon. This statement concerns the physical metric probed by the scalar.

Use the normalized radial coordinate and frequency
\begin{equation} \frac{\dd x}{\dd r}=\frac{R}{\sqrt{k}\Delta_\ell}, \qquad \widetilde\omega=\omega-m\Omega_H, \qquad \frac{\dd^2U}{\dd x^2}+\left[\left(\omega-\frac{m\sqrt{k}a}{R}\right)^2-kV_{Lm}\right]U=0, \label{eq:state-radial-equation} \end{equation}
where $V_{Lm}$ is given in Eq.~\eqref{eq:metric-potential}. Let $Y_{\omega Lm}(\theta,\phi)=S_{Lm}(\theta;a\omega/\sqrt{k})e^{\ii m\phi}/\sqrt{2\pi}$, with $\int|Y_{\omega Lm}|^2\dd\Omega=1$. The radial solutions are fixed by
\begin{equation} U^{\rm in}\sim e^{-\ii\omega x}+\mathcal R_{\rm in}e^{\ii\omega x}\quad(x\to+\infty), \qquad U^{\rm in}\sim\mathcal T_{\rm in}e^{-\ii\widetilde\omega x}\quad(x\to-\infty), \label{eq:state-in-boundary-data} \end{equation}
and
\begin{equation} U^{\rm up}\sim e^{\ii\widetilde\omega x}+\mathcal R_{\rm up}e^{-\ii\widetilde\omega x}\quad(x\to-\infty), \qquad U^{\rm up}\sim\mathcal T_{\rm up}e^{\ii\omega x}\quad(x\to+\infty). \label{eq:state-up-boundary-data} \end{equation}
The corresponding modes and Klein--Gordon product are
\begin{equation} f^{\rm in}_{\omega Lm}=\frac{e^{-\ii\omega\tau}Y_{\omega Lm}U^{\rm in}}{\sqrt{4\pi\omega R}}, \qquad f^{\rm up}_{\omega Lm}=\frac{e^{-\ii\omega\tau}Y_{\omega Lm}U^{\rm up}}{\sqrt{4\pi\nu R}}, \qquad \nu=\widetilde\omega>0, \label{eq:state-normalized-modes} \end{equation}
\begin{equation} (f,g)_{\rm KG}=\ii\int_{\mathcal C}\dd\Sigma^\mu\left[f^*\partial_\mu g-(\partial_\mu f)^*g\right]. \label{eq:state-KG-product} \end{equation}
Here, the in family has $\omega>0$, while the up family has positive horizon frequency $\nu>0$ and $\omega=\nu+m\Omega_H$. In particular, the up family includes negative Killing frequencies. On the past boundaries, the in modes have data $e^{-\ii\omega(\tau+x)}Y/\sqrt{4\pi\omega R}$ at infinity and zero horizon data; the up modes have the opposite support. With an angle $\varphi_H$ transported along the horizon generators, their horizon data are $e^{-\ii\nu u}Y_{\omega Lm}(\theta,\varphi_H)/\sqrt{4\pi\nu R_+}$, where $u=\tau-x$ is the limiting horizon parameter. Evaluating Eq.~\eqref{eq:state-KG-product} on these past boundaries gives
\begin{equation} (f^A_{\xi Lm},f^B_{\xi'L'm'})_{\rm KG}=\delta^{AB}\delta(\xi-\xi')\delta_{LL'}\delta_{mm'}, \qquad \xi=\omega\ \hbox{for in}, \qquad \xi=\nu\ \hbox{for up}. \label{eq:state-mode-normalization} \end{equation}
Complex conjugate modes have the opposite norm. Fourier transformation and the angular basis span the prescribed incoming radiative data. We use their propagated solutions as the radiative sector. For $\ell=0$, asymptotic completeness is established by Kerr scattering theory~\cite{DafermosEtAl2018Scattering}. For $\ell\neq0$, completeness for all bulk solutions and the Hadamard property of the resulting state are not established here. Additionally, the construction below specifies the state and frequency map on the stated radiative sector.


\subsubsection{Affine frequency transformation and the state}

The regular metric horizon supplies the affine coordinate $U=-\kappa^{-1}e^{-\kappa u}$ on the past horizon, with $\chi=-\kappa\, U\partial_U$ and $\chi=\partial_\tau+\Omega_H\partial_\phi$. Continue this characteristic surface through its bifurcation point to $U>0$. After absorbing $\sqrt{R_+}$ into the horizon data and separating a normalized angular function, the reduced null product is $\ii\int\dd U\,f^*\overleftrightarrow{\partial_U}g$. Its positive norm Killing modes and positive affine frequency modes are
\begin{equation} p^R_\nu=\frac{\Theta(-U)(-\kappa U)^{\ii\nu/\kappa}}{\sqrt{4\pi\nu}}, \qquad p^L_\nu=\frac{\Theta(U)(\kappa U)^{-\ii\nu/\kappa}}{\sqrt{4\pi\nu}}, \qquad F_\Omega=\frac{e^{-\ii\Omega U}}{\sqrt{4\pi\Omega}}, \qquad \nu,\Omega>0. \label{eq:state-horizon-bases} \end{equation}
The superscripts $R,L$ label the two portions of this initial characteristic surface. The affine data define the vacuum on the extended characteristic line. Their restrictions to $U<0$ are propagated into the right exterior with zero data at past infinity. The affine coordinate is used on the characteristic surface; on the other hand, notice that $u$ is not assumed to be an exact null optical function throughout the rotating exterior.

Writing $p^R_\nu=\int_0^\infty\dd\Omega\,[\alpha_{\nu\Omega}F_\Omega+\beta_{\nu\Omega}F_\Omega^*]$, the regulated Fourier integral gives
\begin{equation} 
\int_0^\infty\dd s\,(\kappa s)^{\ii\nu/\kappa}e^{-(\epsilon+\ii\Omega)s}=\frac{\kappa^{\ii\nu/\kappa}\Gamma(1+\ii\nu/\kappa)}{(\epsilon+\ii\Omega)^{1+\ii\nu/\kappa}}, \qquad \epsilon\downarrow0. \label{eq:state-Mellin-integral} 
\end{equation}
{In this phase convention, the coefficients are}
\begin{equation} \alpha_{\nu\Omega}=\frac{1}{2\pi\kappa}\sqrt{\frac{\nu}{\Omega}}e^{\pi\nu/(2\kappa)}\Gamma\!\left(\frac{\ii\nu}{\kappa}\right)\left(\frac{\kappa}{\Omega}\right)^{\ii\nu/\kappa}, \qquad \beta_{\nu\Omega}=-e^{-\pi\nu/\kappa}\alpha_{\nu\Omega}. \label{eq:state-Bogoliubov-kernels} \end{equation}
This calculation fixes the frequency transformation from the chosen affine modes~\cite{ZilbermanEtAl2022}. Its normalization follows from $|\Gamma(\ii y)|^2=\pi/[y\sinh(\pi y)]$ and the Fourier delta function in $\ln\Omega$. With $N_\nu=(e^{2\pi\nu/\kappa}-1)^{-1}$, we obtain
\begin{equation} \int_0^\infty\dd\Omega\left(\alpha_{\nu\Omega}\alpha_{\nu'\Omega}^*-\beta_{\nu\Omega}\beta_{\nu'\Omega}^*\right)=\delta(\nu-\nu'), \qquad \int_0^\infty\dd\Omega\,\beta_{\nu\Omega}\beta_{\nu'\Omega}^*=N_\nu\delta(\nu-\nu'). \label{eq:state-kernel-normalization} \end{equation}
For the full angular dependence, let $Y_A$ be a fixed orthonormal basis on the horizon and $Y_c(\nu)$ the spheroidal harmonic of a Killing mode, with $c=(L,m)$. The coefficients include the angular overlaps
\begin{equation} C_{Ac}(\nu)=\int\dd\Omega_2\,Y_A^*Y_c(\nu), \qquad D_{Ac}(\nu)=\int\dd\Omega_2\,Y_A Y_c(\nu). \label{eq:state-angular-overlaps} \end{equation}
The four dimensional kernels are given by
\begin{equation} \alpha_{\nu c,\Omega A}=\alpha_{\nu\Omega}C_{Ac}(\nu), \qquad \beta_{\nu c,\Omega A}=\beta_{\nu\Omega}D_{Ac}(\nu). \label{eq:state-angular-kernels} \end{equation}
Completeness of $Y_A$ restores the angular Kronecker deltas in Eq.~\eqref{eq:state-kernel-normalization}. In this way, the angular conversion is included even though the spheroidal functions depend on frequency.

The state is the Gaussian vacuum annihilated by the in operators at past infinity and the affine up operators. If $F^{\rm up}_{\Omega A}(X)$ denotes the propagated restriction of the affine data to $U<0$, its radiative two point function is
\begin{equation} W_U(X,X')=\sum_{Lm}\int_0^\infty\dd\omega\,f^{\rm in}_{\omega Lm}(X)f^{{\rm in}*}_{\omega Lm}(X')+\sum_A\int_0^\infty\dd\Omega\,F^{\rm up}_{\Omega A}(X)F^{{\rm up}*}_{\Omega A}(X'). \label{eq:state-Unruh-covariance} \end{equation}
The Killing up sector therefore has occupation $N_\nu$, while the incoming infinity sector is empty. This is an Unruh prescription with no incoming thermal bath. In the Kerr limit, the Kay--Wald obstruction to a regular symmetry preserving scalar Hartle--Hawking state remains in force~\cite{KayWald1991}; deleting superradiant modes from a thermal product does not change that result.


\subsubsection{Scattering map and superradiant channels}

The remaining coefficients follow from the exact radial equation. Let $Q(x)$ denote the coefficient multiplying $U$ in Eq.~\eqref{eq:state-radial-equation}. The Jost solutions are specified by the convergent integral equations
\begin{equation} j_+(x)=e^{\ii\omega x}+\int_x^\infty\frac{\sin[\omega(y-x)]}{\omega}[\omega^2-Q(y)]j_+(y)\,\dd y, \label{eq:state-Jost-infinity} \end{equation}
\begin{equation} h_-(x)=e^{-\ii\widetilde\omega x}+\int_{-\infty}^x\frac{\sin[\widetilde\omega(x-y)]}{\widetilde\omega}[\widetilde\omega^2-Q(y)]h_-(y)\,\dd y. \label{eq:state-Jost-horizon} \end{equation}
The two potential differences decay as $x^{-2}$ at infinity and exponentially at the horizon, respectively. Put $j_-=j_+^*$, $h_+=h_-^*$, $W(f,g)=f\partial_xg-(\partial_xf)g$, $\mathcal D=W(h_-,j_+)$, and $s=\operatorname{sgn}\widetilde\omega$. For $\omega>0$, $\widetilde\omega\neq0$, and $\mathcal D\neq0$, the coefficients are
\begin{equation} \mathcal R_{\rm in}=-\frac{W(h_-,j_-)}{\mathcal D}, \qquad \mathcal R_{\rm up}=-\frac{W(h_+,j_+)}{\mathcal D}, \qquad \mathcal T_{\rm in}=\frac{2\ii\omega}{\mathcal D}, \qquad \mathcal T_{\rm up}=\frac{2\ii\widetilde\omega}{\mathcal D}. \label{eq:state-Jost-coefficients} \end{equation}
These expressions keep the scattering phases. With $t=2\ii\sqrt{\omega|\widetilde\omega|}/\mathcal D$, the complete two channel map at each radiative frequency and its conserved flux relation are shown below
\begin{equation} S=\begin{pmatrix}\mathcal R_{\rm in}&s\,t\\t&\mathcal R_{\rm up}\end{pmatrix}, \qquad S^\dagger\eta_s S=\eta_s, \qquad \eta_s=\operatorname{diag}(1,s), \qquad \Gamma=1-|\mathcal R_{\rm in}|^2=\frac{\widetilde\omega}{\omega}|\mathcal T_{\rm in}|^2=s|t|^2. \label{eq:state-scattering-matrix} \end{equation}
Here, $\Gamma$ is the signed absorption probability $\mathcal A^{(\met)}$ used above. The frequency transformation in Eqs.~\eqref{eq:state-Bogoliubov-kernels}--\eqref{eq:state-angular-kernels}, followed by $S$, specifies the radiative Bogoliubov map. The Jost functions are determined by Eq.~\eqref{eq:state-radial-equation}; in addition, notice that the horizon Fourier calculation supplies no substitute for this scattering step.

For $\widetilde\omega>0$, the canonical relation is $(a_{\rm out},b_{\rm down})^T=S(a_{\rm in},b_{\rm up})^T$. For $\widetilde\omega<0$, the positive horizon frequency is $\nu=-\widetilde\omega$ in the conjugate channel $(-\omega,L,-m)$, and the relation becomes $(a_{\rm out},b_{\rm down}^\dagger)^T=S(a_{\rm in},b_{\rm up}^\dagger)^T$. In particular,
\begin{equation} a_{\rm out}=\mathcal R_{\rm in}a_{\rm in}+t\,b_{\rm up}\quad(\widetilde\omega>0), \qquad a_{\rm out}=\mathcal R_{\rm in}a_{\rm in}-t\,b_{\rm up}^\dagger\quad(\widetilde\omega<0). \label{eq:state-superradiant-operator} \end{equation}
The horizon operator in this relation is expressed in the affine vacuum basis by
\begin{equation} b_{\nu c}=\sum_A\int_0^\infty\dd\Omega\left[\alpha_{\nu c,\Omega A}^*c_{\Omega A}-\beta_{\nu c,\Omega A}^*c_{\Omega A}^\dagger\right]. \label{eq:state-composed-operator-kernel} \end{equation}
Together these equations give the complex annihilation and creation kernels of the outgoing radiative modes.
Using the vacuum in data and $\langle b_{\rm up}^\dagger b_{\rm up}\rangle=N_{|\widetilde\omega|}$ gives
\begin{equation} \langle n\rangle_{\rm out}=\Gamma N_{\widetilde\omega}\quad(\widetilde\omega>0), \qquad \langle n\rangle_{\rm out}=-\Gamma(1+N_{-\widetilde\omega})\quad(\widetilde\omega<0), \qquad \langle n\rangle_{\rm out}=\frac{\Gamma}{e^{\widetilde\omega/T_{\met}}-1}. \label{eq:state-derived-Page-factor} \end{equation}
The superradiant term includes spontaneous emission and has a positive mean occupation. No superradiant channel is omitted. The threshold is understood through the joint limit of the signed absorption factor and the denominator.

As a numerical scalar example, for $M=0.2$, $a=0.1$, $k=1.15$, and $L=m=1$, the physical angular velocity is $\Omega_H=0.73437$. Integrating Eq.~\eqref{eq:state-radial-equation} at $\omega=0.4,0.5$ gives $\Gamma=(-4.23,-6.88)\times10^{-5}$ and $\langle n\rangle_{\rm out}=(4.93,9.24)\times10^{-5}$. These values include the superradiant creation term. We integrate from $r-r_+=10^{-9}$ to $r=1600$, using the regular spheroidal eigenvalue and relative tolerance $10^{-12}$. Increasing the outer fitting radius from $800$ to $1600$ changes $\Gamma$ by less than $4.6\times10^{-10}$; the absolute flux identity error is below $1.5\times10^{-14}$.


\subsubsection{Horizon correlations}

The affine modes also determine horizon pair correlations on the extended characteristic surface. Their oscillator representation and packet covariance are collected in Appendix~\ref{app:horizon-pairs}. These correlations follow from the specified state and supply no general factorization of the full exterior and interior fields into independent pairs.


\subsubsection{Boundary data at the singular axes}

For the rotating \textit{metric--affine} geometry, the axial boundary condition can be calculated directly. On either open exterior axis, let $\vartheta=\theta$ or $\pi-\theta$, $D=r^2+a^2$, $P_0=\alpha r^2+\beta a^2$, and $H_{\rm ax}=qD^2/(\beta P_0)$. Including $g_{r\theta}$, meridional normal coordinates have the leading relations
\begin{equation} \zeta=r+\frac{rXa\sqrt{\Delta}}{P_0}\vartheta+O(\vartheta^2), \qquad \rho=\sqrt{H_{\rm ax}(\zeta)}\,\vartheta+O(\vartheta^2), \qquad \dd z=\sqrt{\frac{P_0}{q\beta\Delta}}\,\dd\zeta. \label{eq:ma-axis-normal-coordinates} \end{equation}
The coefficients are evaluated at the footpoint $\zeta(z)$. Extending these as Gaussian normal coordinates gives $g_{\rho\rho}=1$, $g_{\rho z}=0$, $g_{zz}=1+O(\rho)$, and
\begin{equation} g_{\phi\phi}=c(z)^2\rho^2[1+O(\rho)], \qquad \sqrt{-g}=N_0c\rho[1+O(\rho)], \qquad c^2=\frac{P_0}{\alpha D}, \qquad N_0^2=\frac{\Delta}{D}. \label{eq:ma-axis-measure} \end{equation}
Here, the determinant uses $\tau_{\MA}$. For $\Phi=e^{-\ii\omega\tau_{\MA}+\ii m\phi}\psi_m(z,\rho)$, the leading transverse operator in Eq.~\eqref{eq:scalar-operator} is
\begin{equation} 0=\left[\partial_\rho^2+\frac{1}{\rho}\partial_\rho-\frac{m^2}{c(z)^2\rho^2}\right]\psi_m+O(1)\partial_\rho\psi_m+O(\rho^{-1})\psi_m+\mathcal L_{z,\omega,m}\psi_m. \label{eq:ma-axis-indicial-operator} \end{equation}
The tangential operator has bounded coefficients on a compact open axial segment. It gives lower order logarithmic corrections when it differentiates the position dependent exponents. The indicial equation and normal behaviors are
\begin{equation} p^2=\frac{m^2}{c(z)^2}, \qquad \nu_m=\frac{|m|}{c(z)}, \qquad \psi_m\sim A_m(z)(\rho/\rho_0)^{\nu_m}+B_m(z)(\rho/\rho_0)^{-\nu_m}\quad(m\neq0), \label{eq:ma-axis-indicial-powers} \end{equation}
with $\psi_0=A_0(z)+B_0(z)\ln(\rho/\rho_0)+o(1)$ and a fixed length $\rho_0>0$. The conserved Klein--Gordon current gives the axial boundary form
\begin{equation} \mathcal F_{\rm ax}(\Phi_1,\Phi_2)=\ii\lim_{\epsilon\to0}\int_{\rho=\epsilon}\sqrt{-g}\left[\Phi_1^*\partial_\rho\Phi_2-(\partial_\rho\Phi_1)^*\Phi_2\right]\dd\tau_{\MA}\,\dd z\,\dd\phi. \label{eq:ma-axis-boundary-flux} \end{equation}
The negative power contributes a term proportional to $|B_m|^2\int_0^\epsilon\rho^{-2\nu_m-1}\dd\rho$ to the transverse gradient energy, while the logarithm contributes $|B_0|^2\int_0^\epsilon\dd\rho/\rho$. Both diverge. In this sense, the finite transverse energy selects
\begin{equation} \int_0^\epsilon\rho\left(|\partial_\rho\psi_m|^2+\frac{m^2}{c^2\rho^2}|\psi_m|^2\right)\dd\rho<\infty \quad\Longrightarrow\quad B_m=0, \qquad \lim_{\rho\to0}\rho\partial_\rho\psi_0=0. \label{eq:ma-axis-finite-energy} \end{equation}
We adopt this condition on both excluded axes. It makes their symplectic flux vanish and specifies the axial behavior of the exterior scalar modes. The curvature singularity at $r=r_+$ established in Sec.~II~C supplies no regular affine horizon on which to impose the Unruh data used above. A prescription at that inner boundary and a state remain separate inputs to the coupled exterior problem. Accordingly, the local \textit{metric--affine} contour factor is retained as an exterior propagation result. In the static limit, the extended physical metric in Eq.~\eqref{eq:ma-static-metric} admits the same scalar horizon construction with $\Omega_H=0$ and $\kappa=1/(4M\sqrt H)$, where $H=\sqrt{\alpha}/\beta^{3/2}$; this statement concerns the canonically coupled test field.
}

{\section{Scalar transmission and coupled propagation}
\label{sec:greybody}}

\subsection{Separated equation for the {metric geometry}}
\label{subsec:metric-scattering}

The massless Klein--Gordon equation
\begin{equation}
\Box\Phi=0
\label{eq:KG}
\end{equation}
is separable in Eq.~\eqref{eq:metric-line-element}. With
\begin{equation}
\Phi= \e^{-\ii\bar\omega t+\ii m\phi} S_{Lm}(\theta)R_{Lm}(r),
\label{eq:metric-separation}
\end{equation}
the angular equation is the oblate spheroidal equation,
\begin{equation}
\frac{1}{\sin\theta}\frac{\dd}{\dd\theta} \left(\sin\theta\frac{\dd S_{Lm}}{\dd\theta}\right) +\left[ a^2\bar\omega^2\cos^2\theta -\frac{m^2}{\sin^2\theta} +\Lambda_{Lm} \right]S_{Lm}=0.
\label{eq:spheroidal}
\end{equation}
The radial equation is
\begin{equation}
\frac{\dd}{\dd r} \left(\Delta_{\ell}\frac{\dd R_{Lm}}{\dd r}\right) +\left[ \frac{K^2}{\Delta_{\ell}}-\lambda_{Lm} \right]R_{Lm}=0,
\label{eq:metric-radial}
\end{equation}
where
\begin{equation}
K(r)=\bar\omega(r^2+a^2)-am, \qquad \lambda_{Lm}=\Lambda_{Lm}+a^2\bar\omega^2-2am\bar\omega.
\label{eq:lambda-def}
\end{equation}

\begin{figure}
    \centering
    {\setlength{\fboxsep}{2pt}{\includegraphics[width=0.97\linewidth]{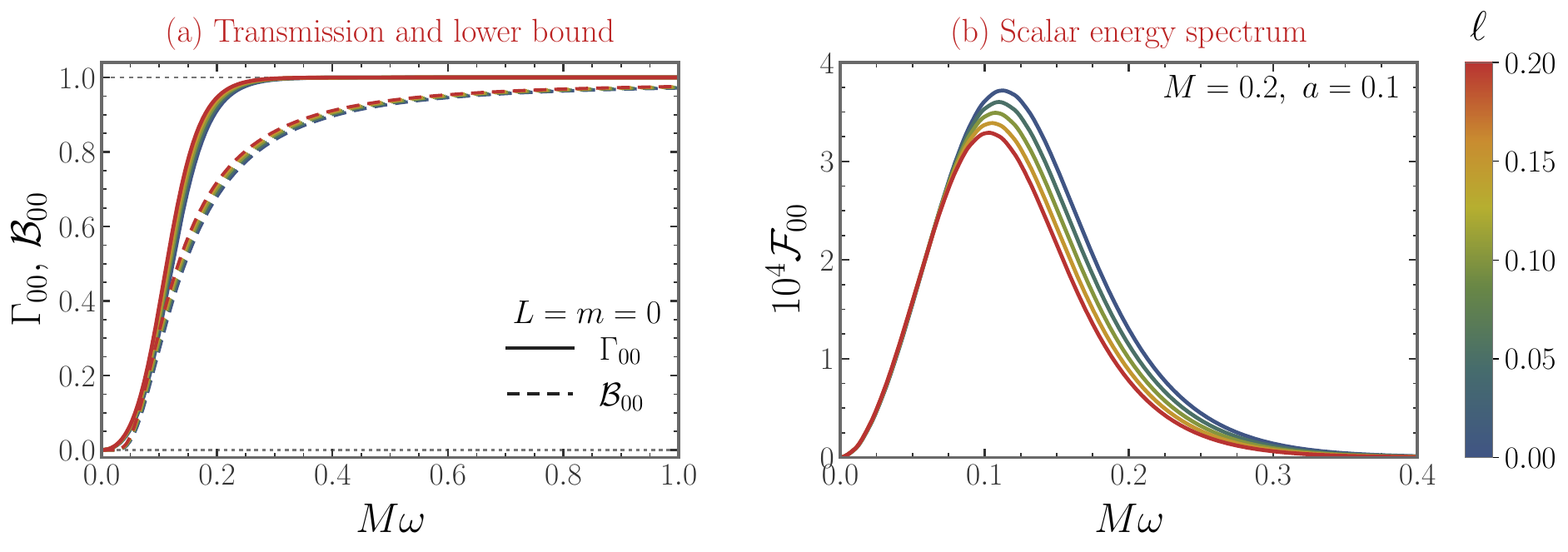}}}
    \caption{{Scalar transmission and partial energy spectrum at identical metric parameters $M=0.2$ and $a=0.1$, for $\ell=0,0.05,0.10,0.15,0.20$. In the left panel, solid curves give the numerical greybody factor $\Gamma_{00}$ and dashed curves give the analytical lower bound $\mathcal B_{00}$, with the spheroidal eigenvalue evaluated at every frequency. The right panel gives the single scalar mode spectrum $\mathcal F_{00}$ defined in Eq.~\eqref{eq:metric-numerical-partial-spectrum}. }}
    \label{greybody}
\end{figure}

Introduce
\begin{equation}
\frac{\dd r_*}{\dd r} =\frac{r^2+a^2}{\Delta_{\ell}}, \qquad U_{Lm}=\sqrt{r^2+a^2}\,R_{Lm}.
\label{eq:metric-tortoise}
\end{equation}
The radial equation becomes
\begin{equation}
\frac{\dd^2U_{Lm}}{\dd r_*^2} +\left[ \left(\bar\omega-\frac{am}{r^2+a^2}\right)^2 -V_{Lm}(r) \right]U_{Lm}=0,
\label{eq:metric-schrodinger}
\end{equation}
with
\begin{equation}
V_{Lm}(r) =\frac{\Delta_{\ell}}{(r^2+a^2)^2} \left[ \lambda_{Lm} +\frac{r\Delta_{\ell}'+\Delta_{\ell}}{r^2+a^2} -\frac{3r^2\Delta_{\ell}}{(r^2+a^2)^2} \right].
\label{eq:metric-potential}
\end{equation}
{

We retain the frequency dependence of the spheroidal eigenvalue. For the axisymmetric modes, $c=a\omega/\sqrt{k}$ and $\lambda_{L0}(c)=\Lambda_{L0}(c)+c^2$ are determined by the regular angular problem
\begin{equation} \left[-\frac1{\sin\theta}\frac{\dd}{\dd\theta}\left(\sin\theta\frac{\dd}{\dd\theta}\right)+c^2\sin^2\theta\right]S_{L0}=\lambda_{L0}(c)S_{L0}, \qquad \lambda_{L0}(c)\geq L(L+1)\geq0. \label{eq:metric-axis-angular-positive} \end{equation}
The inequality follows from the variational ordering of the spherical eigenvalues. For Fig.~\ref{greybody}, we evaluate the oblate characteristic value $\Lambda_{00}(c)$ at every frequency and check it by diagonalizing the angular operator in normalized Legendre functions~\cite{DLMFSpheroidal}. With $y=\cos\theta$, $p_j(y)=\sqrt{(2j+1)/2}\,P_j(y)$, and $b_j=(j+1)/\sqrt{(2j+1)(2j+3)}$, its nonzero matrix entries are
\begin{equation} H_{jj}=j(j+1)+c^2(1-b_j^2-b_{j-1}^2), \qquad H_{j,j+2}=H_{j+2,j}=-c^2b_jb_{j+1}, \qquad b_{-1}=0. \label{eq:metric-angular-spectral-matrix} \end{equation}
The lowest eigenvalue in the even sector gives $\lambda_{00}$. Twelve and twenty four even functions agree with the oblate characteristic value within $10^{-15}$ over the plotted range $0\leq c\leq0.5$.

The radial contribution to the potential also has a definite sign. Set $h=r_+$, $s=r-h$, and $\delta=h^2-ka^2$. Using $2Mh=h^2+ka^2$ and $R=r^2+a^2$ gives
\begin{equation} \frac{r\Delta_\ell'+\Delta_\ell}{R}-\frac{3r^2\Delta_\ell}{R^2}=\frac{P(r)}{kR^2}, \qquad P(h+s)=A_0+A_1s+A_2s^2+2Ms^3. \label{eq:metric-positive-radial-expansion} \end{equation}
Here, $A_0=(h^2+a^2)\delta$, $A_1=\delta(3h^2+2a^2)/h+2(k+1)a^2h$, and $A_2=3h^2+(k+3)a^2$. All coefficients are positive for $M>0$, $k>0$, and $M^2>ka^2$. In this manner, $V_{L0}\geq0$ throughout $r\geq r_+$ for every real frequency, including every curve in Fig.~\ref{greybody}. The integral of $|V_{L0}|$ in the constant comparison function bound equals the integral of $V_{L0}$ in this domain.
}

{

For $m=0$, the transfer matrix inequality~\cite{BoonsermVisser2008,BoonsermVisser2009,GrayVisser2018,AraujoFilho:2024ctw,AraujoFilho:2025rwr,Belchior:2026nnt,Ahmed:2025fwz,Ahmed:2025did,Gogoi:2024epx,Rincon:2024won,LambiaseEtAl2023} gives
\begin{equation} \Gamma_{L0}^{(\met)}(\omega)\geq\mathcal B_{L0}(\omega)\equiv\operatorname{sech}^2\!\left[\frac{\sqrt k\,\mathcal I_{L0}}{2\omega}\right]. \label{eq:metric-greybody-bound} \end{equation}
Here $\Gamma_{L0}$ is the transmission probability, while $\mathcal B_{L0}$ denotes its lower bound. The positive integral $\mathcal I_{L0}=\int_{r_+}^\infty V_{L0}R/\Delta_\ell\,\dd r$ evaluates to
\begin{equation} \mathcal I_{L0}=\frac1a\left[\lambda_{L0}(c)+\frac{k+3}{8k}\right]\tan^{-1}\!\frac a{r_+}-\frac{(k-1)r_+^2+2ka^2}{8kr_+(r_+^2+a^2)}. \label{eq:metric-I} \end{equation}
The value at $a=0$ is understood by continuity.
}
In the static limit,
\begin{equation}
\mathcal{I}_{L0}(a=0) =\frac{\lambda_{L0}}{2M}+\frac{1}{4kM}.
\label{eq:metric-I-static}
\end{equation}
{

For $m\neq0$, the signed absorption and the superradiant creation channel are given by the scattering construction in Sec.~\ref{subsec:quantum-state}. The present bound concerns $m=0$.

We compute the transmission in Fig.~\ref{greybody} by integrating Eq.~\eqref{eq:metric-radial} with unit ingoing horizon data and matching to the asymptotic Jost solutions. The calculation uses $z=M\omega=0.01,0.02,\ldots,1$, $M=0.2$, $a=0.1$, and $\ell=0,0.05,0.10,0.15,0.20$. In units of $M$, we start at $(r-r_+)/M=10^{-7}$, use outer radius $r/M=\max(160,60/z)$, relative tolerance $2\times10^{-11}$, and eight terms in the asymptotic expansion. Decreasing the starting distance by ten, doubling the outer radius, and increasing the asymptotic order to ten changes $\Gamma_{00}$ by less than $4\times10^{-12}$ in the refinement checks. The reflection and transmission flux expressions agree within $8\times10^{-11}$. We report a conservative absolute numerical accuracy of $10^{-9}$; the computed transmission exceeds the bound at all 500 sampled points.

For the scalar radiative Unruh prescription of Sec.~\ref{subsec:quantum-state}, these transmission factors give the partial energy spectrum
\begin{equation} \mathcal F_{00}(z)\equiv M\frac{\dd^2E_{00}}{\dd\tau_{\met}\,\dd\omega}=\frac{z\,\Gamma_{00}(z)}{2\pi\{\exp[z/(MT_{\met})]-1\}}. \label{eq:metric-numerical-partial-spectrum} \end{equation}
For $\ell=0$ and $0.20$, its peak occurs at $z=0.11247$ and $0.10298$, with heights $3.7176\times10^{-4}$ and $3.2893\times10^{-4}$, respectively. The spectra cross at low frequency. It is important to point out that these results describe the single $L=m=0$ scalar channel at the stated metric parameters.
}

\subsection{Mode coupling in the rotating \textit{metric--affine} geometry}
\label{subsec:ma-scattering}

For the \textit{metric--affine} solution, the inverse $(t,\phi)$ block is $q$ times the Kerr inverse block. After setting
\begin{equation}
\Phi=\e^{-\ii\bar\omega t+\ii m\phi}\Psi_m(r,\theta),
\label{eq:ma-wave-ansatz}
\end{equation}
the exact scalar equation is
\begin{align}
0={}& \frac{1}{\sqrt{-g}}\partial_r \left[ \sqrt{-g}\left( g^{rr}\partial_r\Psi_m +g^{r\theta}\partial_\theta\Psi_m \right) \right] \nonumber\\ &+ \frac{1}{\sqrt{-g}}\partial_\theta \left[ \sqrt{-g}\left( g^{r\theta}\partial_r\Psi_m +g^{\theta\theta}\partial_\theta\Psi_m \right) \right] \nonumber\\ &+ \left( -g^{tt}\bar\omega^2 +2g^{t\phi}\bar\omega m -g^{\phi\phi}m^2 \right)\Psi_m .
\label{eq:ma-KG-PDE}
\end{align}
{The mixed meridional derivatives and the joint dependence of their coefficients on $r,\theta$ couple the angular channels for generic $aX\neq0$. We write the reference expansion}
\begin{equation}
\Psi_m(r,\theta) =\sum_L u_{Lm}(r)Y_{Lm}(\theta)
\label{eq:ma-mode-expansion}
\end{equation}
{The functions $Y_{Lm}$ serve as a reference family. Their admissible use follows from the finite energy form below. The derivative coupling begins at order $aX$, while the WKB scale anisotropy in Eq.~\eqref{eq:ma-temperature-expansion} begins at order $Xa^2$.}

{
The finite energy axial condition in Eq.~\eqref{eq:ma-axis-finite-energy} gives a definite operator realization on each compact radial annulus. Fix $r_+<r_0<r_1<\infty$, real $\bar\omega$, and $m\in\mathbb Z$, using Eq.~\eqref{eq:ma-wave-ansatz}. Set $\mathcal D=(r_0,r_1)\times(0,\pi)$ and $\mathcal H_m=L^2(\mathcal D,w\,\dd r\,\dd\theta)$, with $w=\Sigma\sin\theta/(\sqrt\alpha\,\beta^{3/2})$. The matrix $G=(g^{AB})_{A,B=r,\theta}$ has positive diagonal entries and determinant $\beta^2\Delta/\Sigma^2>0$. Its continuity makes it uniformly positive on the closed annulus, with the mixed component retained. Taking all these features together, the potential reads
\begin{equation} 
V_{\bar\omega m}=\frac{qm^2}{\Sigma\sin^2\theta}+V^{\rm b}_{\bar\omega m}, \qquad V^{\rm b}_{\bar\omega m}=g^{tt}\bar\omega^2-2g^{t\phi}\bar\omega m-\frac{qa^2m^2}{\Sigma\Delta}, \qquad C_{\bar\omega m}=\|V^{\rm b}_{\bar\omega m}\|_\infty<\infty. 
\label{eq:ma-annulus-potential} 
\end{equation}
Here, $V_{\bar\omega m}=g^{tt}\bar\omega^2-2g^{t\phi}\bar\omega m+g^{\phi\phi}m^2$. The bounded real term $V^{\rm b}_{\bar\omega m}$ retains the ergoregion contribution. On compactly supported smooth functions, define
\begin{equation} \mathfrak a_{\bar\omega m}[u,v]=\int_{\mathcal D}w\left[g^{AB}(\partial_Au)^*\partial_Bv+\left(\frac{qm^2}{\Sigma\sin^2\theta}+V^{\rm b}_{\bar\omega m}\right)u^*v\right]\dd r\,\dd\theta. \label{eq:ma-annulus-form} \end{equation}
Its finite energy domain is fixed by
\begin{equation} \|u\|_{\mathcal V_m}^2=\int_{\mathcal D}w\left[g^{AB}(\partial_Au)^*\partial_Bu+\left(1+\frac{qm^2}{\Sigma\sin^2\theta}\right)|u|^2\right]\dd r\,\dd\theta, \qquad \mathcal V_m=\overline{C_c^\infty(\mathcal D)}^{\|\cdot\|_{\mathcal V_m}}. \label{eq:ma-annulus-form-domain} \end{equation}
Since $w$ is comparable to $\sin\theta$, $G$ is uniformly positive, and $q/\Sigma$ has positive upper and lower bounds, this norm is equivalent to the usual $H^1$ norm in the fixed $m$ sector of the reference product $(r_0,r_1)\times S^2$. This comparison makes no smoothness assumption on the physical axes. The closure implements the finite energy pole condition in Eq.~\eqref{eq:ma-axis-finite-energy} and Dirichlet radial data.

The form is densely defined, closed, and bounded below by $-C_{\bar\omega m}\|u\|_{\mathcal H_m}^2$. Its Friedrichs realization is the unique self adjoint operator associated with this form and domain~\cite{Kato1995}, with differential expression
\begin{equation} \mathsf A_{\bar\omega m}u=-\frac1w\partial_A\!\left(wg^{AB}\partial_Bu\right)+V_{\bar\omega m}u, \qquad \mathsf A_{\bar\omega m}\geq-C_{\bar\omega m}\,\mathsf I. \label{eq:ma-annulus-Friedrichs} \end{equation}
The reference norm comparison gives a compact embedding into $\mathcal H_m$ so that compact resolvent and a complete orthonormal eigenbasis for this annulus problem. Finite spherical harmonic sums with compactly supported smooth radial coefficients are dense in the form domain, giving Eq.~\eqref{eq:ma-mode-expansion} a variational meaning with the physical weight and its Gram matrix retained. Also, the local radial contour and the annulus spectrum keep their respective exterior meanings.
}

An analytical check is available in the static limit. Define $H=\sqrt{\alpha}/\beta^{3/2}$. After normalizing the time, the static metric is
\begin{equation}
\dd s_{\MA}^{2}\big|_{a=0} =-f\dd\tau_{\MA}^{2} +\frac{H}{f}\dd r^{2} +\frac{r^2}{q}\dd\Omega^{2}.
\label{eq:ma-static-metric}
\end{equation}
The scalar potential is
\begin{equation}
V_{L}^{(\MA,0)} =f\left[ \frac{qL(L+1)}{r^2} +\frac{f'}{Hr} \right], \qquad \frac{\dd r_*}{\dd r}=\frac{\sqrt H}{f}.
\label{eq:ma-static-potential}
\end{equation}
The corresponding lower bound is
\begin{align}
\mathcal{T}_{L}^{(\MA,0)}(\omega) &\geq \operatorname{sech}^{2} \left[ \frac{\mathcal{I}_{L}^{(\MA,0)}}{2\omega} \right], \label{eq:ma-static-bound}\\ \mathcal{I}_{L}^{(\MA,0)} &= \frac{\alpha^{3/4}}{\beta^{1/4}} \frac{L(L+1)}{2M} +\frac{\beta^{3/4}}{\alpha^{1/4}} \frac{1}{4M}.
\label{eq:ma-static-I}
\end{align}
At $X=0$, the Schwarzschild result is recovered.


{\section{Mode fluxes and static emission}
\label{sec:evaporation}}

\subsection{Mode-resolved fluxes}

{Under the conditional {Unruh type} prescription of Eq.~\eqref{eq:metric-page-occupation}, with normalized asymptotic scattering modes, the standard particle, energy, and angular momentum fluxes of species $i$, written with signed absorption probabilities $\Gamma_{Lm}^{(i)}$, are}
\begin{align}
\frac{\dd^2N_i}{\dd{\tau}\,\dd\omega} &= \frac{g_i}{2\pi} \sum_{L,m} \frac{\Gamma_{Lm}^{(i)}(\omega)} {\exp[(\omega-m\Omega_H)/T_H]-(-1)^{2s_i}}, \label{eq:number-flux}\\ \frac{\dd^2E_i}{\dd{\tau}\,\dd\omega} &= \frac{g_i\omega}{2\pi} \sum_{L,m} \frac{\Gamma_{Lm}^{(i)}(\omega)} {\exp[(\omega-m\Omega_H)/T_H]-(-1)^{2s_i}}, \label{eq:energy-flux}\\ \frac{\dd^2J_i}{\dd{\tau}\,\dd\omega} &= \frac{g_i}{2\pi} \sum_{L,m} \frac{m\,\Gamma_{Lm}^{(i)}(\omega)} {\exp[(\omega-m\Omega_H)/T_H]-(-1)^{2s_i}}.
\label{eq:angular-flux}
\end{align}

{

For established canonical charges and outgoing fluxes, the equations are
\begin{equation}
-\frac{\dd E}{\dd\tau}=\sum_i\int_0^\infty\frac{\dd^2E_i}{\dd\tau\,\dd\omega}\,\dd\omega, \qquad -\frac{\dd J}{\dd\tau}=\sum_i\int_0^\infty\frac{\dd^2J_i}{\dd\tau\,\dd\omega}\,\dd\omega.
\label{eq:balance}
\end{equation}
}

{For the rotating \textit{metric--affine} geometry, the radial WKB scale belongs to the nonpolar exterior calculation of Sec.~V. The singular boundary and axes established in Sec.~II~C preclude its use as the global temperature in these flux equations.}

\begin{figure}
    \centering
    \includegraphics[scale=0.55]{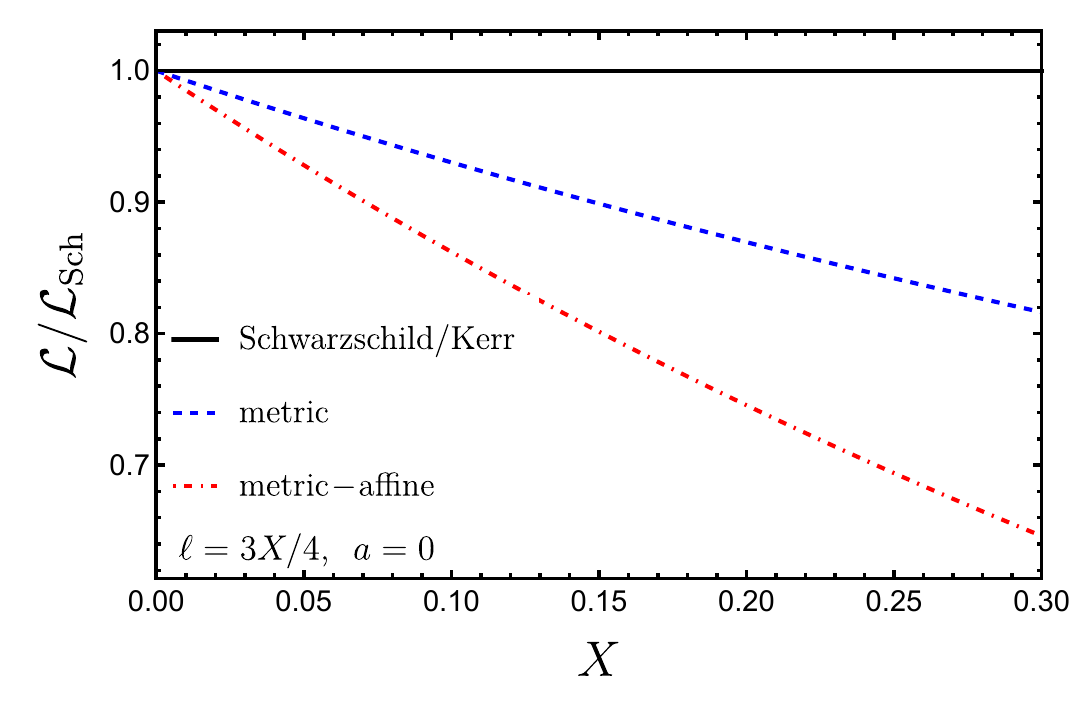}
    \caption{{Static Stefan--Boltzmann luminosity ratios at equal canonical energy, with $\ell=3X/4$. The curves are $1$, $1/(1+\ell)$, and $\beta^3/\alpha$ for Schwarzschild, metric, and \textit{metric--affine} geometries, respectively.}}
    \label{f1}
\end{figure}
\begin{figure}
    \centering
    \includegraphics[scale=0.55]{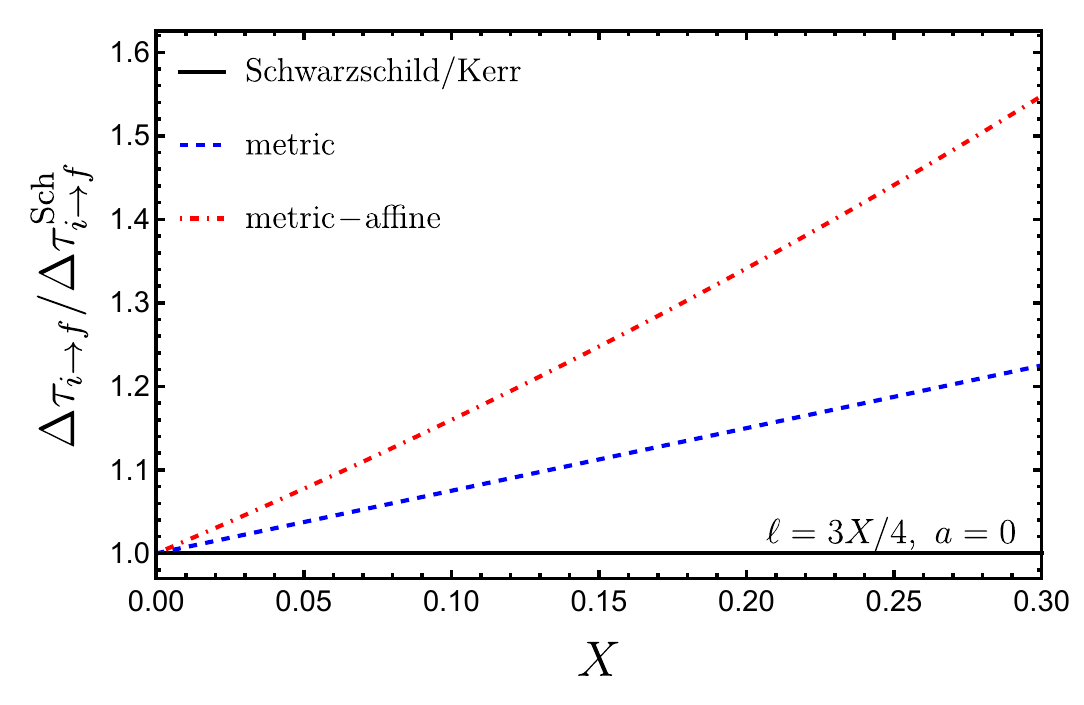}
    \caption{ {Static evolution times normalized to the Schwarzschild value, for common initial and final canonical energies. With $\mathrm{d}E/\mathrm{d}\tau = -\mathcal{L}_{\rm SB}$ and $\ell = 3X/4$, the ratios are $1$, $1 + \ell$, and $\alpha/\beta^3$ for Schwarzschild, metric, and \textit{metric--affine} geometries, respectively.}}
    \label{f2}
\end{figure}

{

\subsection{Static test field emission}
\label{subsec:static-emission}

Assuming a thermal outgoing distribution for canonically coupled test fields on the extended static physical metrics, we use the Stefan--Boltzmann approximation
\begin{equation} \mathcal L_{\mathrm{SB}}=\frac{\pi^2}{120}g_\star\epsilon_{\mathrm{em}}A_HT_H^4, \qquad \gamma=\frac{g_\star\epsilon_{\mathrm{em}}}{30720\pi}, \label{eq:SB} \end{equation}
where $\epsilon_{\mathrm{em}}$ is a {frequency averaged} emissivity. With identical particle content and emissivity, the static luminosities at equal canonical energy are
\begin{equation} \mathcal L_{\met}(E)=\frac{\gamma}{kE^2}, \qquad \mathcal L_{\MA}(E)=\frac{\gamma\beta^3}{\alpha E^2}. \label{eq:ma-static-luminosity} \end{equation}
Assuming the {test field} energy balance $\dd E/\dd\tau=-\mathcal L_{\mathrm{SB}}$, the durations between prescribed canonical energies are
\begin{equation} \Delta\tau_{\met}=\frac{k(E_i^3-E_f^3)}{3\gamma}, \qquad \Delta\tau_{\MA}=\frac{\alpha(E_i^3-E_f^3)}{3\gamma\beta^3}. \label{eq:ma-static-lifetime} \end{equation}
Notice that these estimates follow the specified emission model, independently of a gravitational entropy or {first law} identification.

Figures~\ref{f1} and \ref{f2} compare the static luminosities and model durations at equal canonical energy and common energy endpoints. For $\ell=3X/4$, identical particle content, and identical emissivity, $\mathcal L_{\mathrm{Sch}}>\mathcal L_{\met}>\mathcal L_{\MA}$ and $\Delta\tau_{\mathrm{Sch}}<\Delta\tau_{\met}<\Delta\tau_{\MA}$.

Furthermore, Appendix~\ref{app:geometrical-integrals} collects the rotating geometrical functionals, their expansions, and reciprocal parameter integrals illustrated in Figs.~\ref{f3} and \ref{f4}.
}

{

\section{Parameter domains and static observational estimates}
\label{sec:bounds}
}

{

The observational estimates in this section concern $a=0$, geodesic test matter minimally coupled to the physical metric, the normalized asymptotic clocks, and the fixed azimuthal period of Sec.~\ref{subsec:charges}. Applying a Solar System estimate to another source assumes the same constant coupling, vacuum norm, and static vacuum profile, with the source mass expressed in the same convention. The nonzero deformation values in the figures illustrate parameter dependence; for example, $\ell=0.15$ exceeds the conditional static Cassini sensitivity quoted below by about eleven orders of magnitude.

For the rotating \textit{metric--affine} geometry, geodesic initial data determine unique local rays while they remain in the smooth nonpolar exterior. A global shadow needs a prescription for rays that reach the singular inner boundary or axes. A disk spectrum also needs a source model and radiative transfer to the observer; an asymptotic wave flux needs global boundary data and a state. The scalar operator on bounded annuli in Sec.~\ref{subsec:ma-scattering} supplies a definite stationary problem on those annuli. The present rotating results show their exterior propagation meaning, with no global shadow, disk spectrum, or asymptotic emission prediction assigned to this configuration.
}


\subsection{Admissible parameter domains and the extremal bound}

For the {metric geometry}, the conditions $k=1+\ell>0$ and
\begin{equation}
M^{2}\geq (1+\ell)a^{2}
\label{eq:metric-bound-original}
\end{equation}
are required for the existence of an event horizon. {Introducing the parameter ratio $\chi=a/M$, the metric root condition gives}
\begin{equation}
-1<\ell\leq\frac{1}{\chi^{2}}-1.
\label{eq:metric-ell-bound}
\end{equation}
For the positive deformation considered throughout most of this work, the allowed interval becomes
\begin{equation}
0\leq\ell\leq\ell_{\max}(\chi), \qquad \ell_{\max}(\chi)\equiv\frac{1}{\chi^{2}}-1.
\label{eq:metric-positive-ell-bound}
\end{equation}
{For the \textit{metric--affine} configuration, the canonical charges give the root condition $q^2j_{\rm MA}^2\leq1$. The admissible metric coefficients also require}
\begin{equation}
\alpha=1+\frac{3X}{4}>0, \qquad \beta=1-\frac{X}{4}>0
\label{eq:MA-alpha-beta-bounds}
\end{equation}
{which imply}
\begin{equation}
-\frac{4}{3}<X<4.
\label{eq:MA-X-domain}
\end{equation}
Furthermore,
\begin{equation}
q^{2}=\alpha\beta =1+\frac{X}{2}-\frac{3X^{2}}{16},
\label{eq:q-bound-expression}
\end{equation}
so that the condition $q>1$, used in deriving several of the positive deformation hierarchies, is satisfied for
\begin{equation}
0<X<\frac{8}{3}.
\label{eq:q-positive-bound}
\end{equation}
The calculations based on the weak--deformation expansion require the more restrictive regime $|X|\ll1$.


{

\subsection{Static mass normalization and Solar System estimates}

In areal radius $R$, both static physical metrics take the form
\begin{equation} \dd s^2=-F(R)\dd\tau^2+\frac{\mathcal C_s}{F(R)}\dd R^2+R^2\dd\Omega^2, \qquad F(R)=1-\frac{2\mu}{R}. \label{eq:phenom-static-areal} \end{equation}
Here, $R=r$ and $\mathcal C_s=k$ for the metric geometry, while $R=r/\sqrt q$ and $\mathcal C_s=\alpha/\beta$ for the \textit{metric--affine} geometry. Circular geodesics satisfy $\Omega_{\rm orb}^2=\mu/R^3$ with respect to $\tau$. The orbital mass parameter and canonical energy obey
\begin{equation} \mu_{\rm met}=M, \qquad E_{\rm met}=\sqrt k\,\mu_{\rm met}, \qquad \mu_{\rm MA}=\frac{M}{\sqrt q}, \qquad E_{\rm MA}=q\mu_{\rm MA}. \label{eq:phenom-orbital-mass} \end{equation}
An orbital mass estimate enters the comparison at fixed canonical energy through these relations.

For the metric static vacuum, the normalized metric and deformation parameter coincide with those used in Ref.~\cite{Casana2018}. Applying its Cassini error prescription to the absolute leading residual gives the conditional sensitivity
\begin{equation} |\ell|\lesssim6.2\times10^{-13}. \label{eq:ell-cassini-bound} \end{equation}
This estimate considers the source assumptions about fitted mass, distances, and signal propagation. It is an error based static estimate, with both signs of $\ell$ included.

The static metric used for the reported $X<7.4\times10^{-12}$ estimate in Ref.~\cite{AraujoFilho2023PRD} has areal radial factor $\sqrt\alpha/\beta^{3/2}$. The present static limit has $\mathcal C_s=\alpha/\beta$, so we evaluate its perihelion response directly. For $u=1/R$ and specific angular momentum $\mathcal J$, timelike geodesics give
\begin{equation} \mathcal C_s\frac{\dd^2u}{\dd\phi^2}+u=\frac{\mu}{\mathcal J^2}+3\mu u^2. \label{eq:phenom-Binet} \end{equation}
The variable $\phi/\sqrt{\mathcal C_s}$ reduces this equation to the Schwarzschild orbit equation. At fixed $\mu$ and areal turning radii, the excess advance per radial period is 
\begin{equation} \delta\varpi\equiv\Delta\varpi-\Delta\varpi_{\rm Sch}=\left(\sqrt{\mathcal C_s}-1\right)\left(2\pi+\Delta\varpi_{\rm Sch}\right). \label{eq:phenom-perihelion-excess} \end{equation}
This relation is exact when the Schwarzschild advance is evaluated exactly. For an allowed magnitude $|\delta\varpi|\leq\epsilon_\varpi$, measured in radians per orbit, define
\begin{equation} \eta_\varpi=\frac{\epsilon_\varpi}{2\pi+\Delta\varpi_{\rm Sch}}, \qquad d_\pm=(1\pm\eta_\varpi)^2-1, \qquad 0\leq\eta_\varpi<1. \label{eq:phenom-residual-domain} \end{equation}
The corresponding signed parameter intervals are
\begin{equation} d_-\leq\ell\leq d_+, \qquad \frac{4d_-}{4+d_-}\leq X\leq\frac{4d_+}{4+d_+}. \label{eq:X-mercury-bound} \end{equation}
In the weak field and small deformation limit, $\delta\varpi_{\rm MA}=\pi X+O(X\mu/p,X^2)$, where $p$ is the areal semilatus rectum. Thereby, $|X|\lesssim\epsilon_\varpi/\pi$ is the leading conditional sensitivity. The tolerance $\epsilon_\varpi$ belongs to a fit of orbital and source parameters in the present static geometry. The published numerical estimate for the different static metric is retained only as source context. 


\subsection{Error propagation in the static emission model}

The exact static ratios at equal canonical energy are given in Secs.~IV~B and VII~B. At fixed energy and emission normalization, first order fractional tolerances $\epsilon_T$ and $\epsilon_L$ give
\[
|X|\lesssim4\epsilon_T, \qquad |\ell|\lesssim\epsilon_L, \qquad |X|\lesssim\frac23\epsilon_L.
\]
The temperature supplies no constraint on $\ell$ at fixed $E$. Energy and emissivity uncertainties enter through $T\propto E^{-1}$ and $\mathcal L\propto\gamma E^{-2}$. Their degeneracies with the deformations are explicit in
\[
\delta\ln\mathcal L_{\rm met}=\delta\ln\gamma-2\delta\ln E-\frac{\delta\ell}{k},
\]
\[
\delta\ln\mathcal L_{\rm MA}=\delta\ln\gamma-2\delta\ln E-\frac{3(4+X)}{8\alpha\beta}\,\delta X.
\]
These are error propagation relations within the stated Stefan--Boltzmann model, with common particle content and emissivity. A forecast would also include the complete greybody spectrum, parameter correlations, source and instrument models, and the relevant global boundary conditions. 
}


{

\section{Conclusion}
\label{sec:conclusion}

We compared scalar propagation and static emission estimates in the corrected Newman--Janis metric geometry and the stationary \textit{metric--affine} bumblebee geometry. Their normalized asymptotic clocks fixed the frequencies and angular velocities, while their distinct spatial asymptotics entered the physical charge definitions. At equal canonical energy, the static temperatures satisfied $T_{\met}=T_{\mathrm{Sch}}$ and $T_{\MA}=\beta \,T_{\mathrm{Sch}}$. The illustrative choice $\ell=3X/4$ specified the parameters of these two different geometries.

For generic rotation, the physical metric curvature and trace free nonmetricity diverged at the \textit{metric--affine} outer root. The complete axial coefficient was also positive throughout the open exterior. These results distinguished the singular nonpolar boundary and exterior axes from the independent fixed point obstruction at bifurcation surface poles. The \textit{metric--affine} radial contour factor described local propagation on the punctured exterior. Its finite limiting area and parameter derivatives kept a geometrical meaning.

For the metric geometry, normalized radiative modes and the affine horizon transformation defined an Unruh prescription on the stated radiative sector. The exact radial scattering map included spontaneous superradiant emission. The spheroidal eigenvalues, positivity proof, and converged numerical transmission supported the analytical bound and partial scalar energy spectrum in Fig.~\ref{greybody}. Full bulk completeness and Hadamard regularity for nonzero $\ell$ remained open.

For the rotating \textit{metric--affine} geometry, the mixed meridional component coupled radial and polar propagation. Finite transverse energy fixed the adopted axial behavior and eliminated the axial symplectic flux. The resulting closed form defined a self adjoint stationary operator on bounded exterior annuli. Its extension to the complete exterior, the inner boundary state, and the coupled asymptotic scattering matrix remained separate unresolved elements of the global emission problem. The static emission and Solar System comparisons used the stated test field, mass, and matter assumptions. Their signed conditional intervals and error propagation relations retained this scope.
}

\section*{Acknowledgments}
\hspace{0.5cm} A.A.A.F. is supported by Conselho Nacional de Desenvolvimento Cient\'{\i}fico e Tecnol\'{o}gico (CNPq) project number 150223/2025-0. The author is also indebted to L. A. Cordeiro for fruitful suggestions and discussions, as well as for the encouragement provided throughout the development of this work.

\section*{Data Availability Statement}

{The figures and numerical results were obtained from the equations and parameter choices specified in the manuscript. The corresponding code and numerical data are available from the author upon reasonable request. No external numerical data were used.}

\appendix

{

\section{Static clock normalization}
\label{sec:time-normalization}

For $g_{tt}\to-C$, the normalized clock and its Killing generator obey
\begin{equation} \tau=\sqrt C\,t, \qquad \xi_{\mathrm{phys}}=C^{-1/2}\partial_t, \qquad (T,\omega,\Omega)_{\mathrm{phys}}=C^{-1/2}(T,\omega,\Omega)_t. \label{eq:digression-normalized-killing} \end{equation}
The combination $(\omega-m\Omega)/T$ is invariant. Applying this rescaling with $C_{\met}=1/k$ and $C_{\MA}=1/q$ gives, at equal metric parameter $M$,
\begin{equation} (T_{\met,t},T_{\met})=\frac{(k^{-1},k^{-1/2})}{8\pi M}, \qquad (T_{\MA,t},T_{\MA})=\frac{(\sqrt{\beta/\alpha},\beta^{3/4}/\alpha^{1/4})}{8\pi M}. \label{eq:digression-static-temperatures} \end{equation}
Thus $T_{\met}/T_{\mathrm{Sch}}=1-\ell/2+\order(\ell^2)$ and $T_{\MA}/T_{\mathrm{Sch}}=1-3X/8+\order(X^2)$. The equality of their leading corrections at $\ell=3X/4+\order(X^2)$ is a parameter convention at fixed $M$. This normalization accounts for the difference from the static expressions compared in Ref.~\cite{AraujoFilho2025JCAP}. The comparison at equal canonical energy appears in Sec.~\ref{sec:thermodynamics}.

Within the static Stefan--Boltzmann model of Sec.~\ref{sec:evaporation}, the corresponding ratios at equal $M$ are
\begin{equation} \frac{A_{\MA}}{A_{\mathrm{Sch}}}=\frac1q, \qquad \frac{\mathcal L_{\met}}{\mathcal L_{\mathrm{Sch}}}=\frac1{k^2}, \qquad \frac{\mathcal L_{\MA}}{\mathcal L_{\mathrm{Sch}}}=\frac{\beta^{5/2}}{\alpha^{3/2}}. \label{eq:digression-static-luminosities} \end{equation}
The use of a capture cross section in place of the area specifies a different emission approximation.
}

{

\section{Horizon pair identities}
\label{app:horizon-pairs}

For conjugate angular partners and $z_\nu=e^{-\pi\nu/\kappa}$, the positive affine pair modes and annihilation operators are
\begin{equation} P_\nu=\frac{p^R_\nu+z_\nu p^{L*}_\nu}{\sqrt{1-z_\nu^2}}, \qquad Q_\nu=\frac{p^L_\nu+z_\nu p^{R*}_\nu}{\sqrt{1-z_\nu^2}}. \label{eq:state-analytic-pair-modes} \end{equation}
\begin{equation} c_P=\frac{b_R-z_\nu b_L^\dagger}{\sqrt{1-z_\nu^2}}, \qquad c_Q=\frac{b_L-z_\nu b_R^\dagger}{\sqrt{1-z_\nu^2}}. \label{eq:state-analytic-pair-operators} \end{equation}
Analyticity in the lower complex $U$ half plane fixes these combinations. The vacuum conditions $c_P|0_U\rangle=c_Q|0_U\rangle=0$ give the regulated spectral oscillator representation
\begin{equation} |0_U\rangle_\nu=\sqrt{1-z_\nu^2}\sum_{n=0}^\infty z_\nu^n|n_R,n_L\rangle, \qquad \rho_{R,\nu}=(1-z_\nu^2)\sum_{n=0}^\infty z_\nu^{2n}|n_R\rangle\langle n_R|. \label{eq:state-derived-pair-entropy} \end{equation}
Its entropy is $s_\nu=(1+N_\nu)\ln(1+N_\nu)-N_\nu\ln N_\nu$. Normalized frequency packets supported away from $\nu=0$ have exact covariance
\begin{equation} N_{jk}=\int_0^\infty\dd\nu\,f_j^*(\nu)f_k(\nu)N_\nu, \qquad \int_0^\infty\dd\nu\,f_j^*(\nu)f_k(\nu)=\delta_{jk}. \label{eq:state-packet-covariance} \end{equation}
The central frequency pair approximation applies to narrow packets; exterior scattering also correlates outgoing and reflected channels.
}

{

\section{Geometrical parameter integrals}\label{app:geometrical-integrals}

For the rotating geometries, we retain the area weighted fourth powers of the temperature scales as geometrical functionals $\mathcal P$, with dimensions of power. For the metric geometry,
\begin{equation} \mathcal P_{\met}=\frac{g_\star\epsilon_{\mathrm{em}}}{480\pi}\frac{d_\ell^{\,4}}{k^2\left[\left(r_+^{(\met)}\right)^2+a^2\right]^3}. \label{eq:metric-luminosity-exact} \end{equation}
At fixed $a$ and for slow rotation,
\begin{equation} \mathcal P_{\met}=\frac{\gamma}{k^2M^2}\left[1-\frac{2k+3}{4}\frac{a^2}{M^2}+\order\!\left(\frac{a^4}{M^4}\right)\right]. \label{eq:metric-luminosity-slow} \end{equation}
For the \textit{metric--affine} geometry, the improper exterior limit is
\begin{equation} \mathcal P_{\MA}^{\mathrm{loc}}=\frac{\pi^2}{120}g_\star\epsilon_{\mathrm{em}}\lim_{\epsilon\downarrow0}\int_{\epsilon}^{\pi-\epsilon}\!\int_0^{2\pi}T_{\MA}^4(\theta)\,\dd A_{\mathrm{lim}}, \qquad \dd A_{\mathrm{lim}}=\lim_{r_0\downarrow r_+}\sqrt{g_{\theta\theta}g_{\phi\phi}}\big|_{t=\mathrm{constant},\,r=r_0}\,\dd\theta\,\dd\phi. \label{eq:ma-local-luminosity-def} \end{equation}
Here, $T_{\MA}(\theta)$ is the exterior WKB scale and $\dd A_{\mathrm{lim}}$ is the limiting area measure of Sec.~\ref{subsec:areas}. The integral has no luminosity interpretation at infinity. To first order in $X$, with the complete Kerr spin dependence,
\begin{equation} \mathcal P_{\MA}^{\mathrm{loc}}=\mathcal P_{\Kerr}\left[1+X\left(\frac34-\frac{5r_+}{2a}\tan^{-1}\frac a{r_+}\right)\right]+\order(X^2), \label{eq:ma-local-luminosity} \end{equation}
\begin{equation} \mathcal P_{\Kerr}=\frac{g_\star\epsilon_{\mathrm{em}}}{480\pi}\frac{d^4}{(r_+^2+a^2)^3}. \label{eq:kerr-luminosity} \end{equation}
For $\chi=a/M\ll1$ and $|X|\ll1$,
\begin{equation} \mathcal P_{\MA}^{\mathrm{loc}}=\frac{\gamma}{M^2}\left[1-\frac54\chi^2-\frac74X+\frac{115}{48}X\chi^2\right]+\order(\chi^4,X^2,X\chi^4). \label{eq:ma-local-luminosity-slow} \end{equation}

The reciprocal integrals follow prescribed paths in the metric parameters,
\begin{equation} \mathcal I_{i\to f}=\int_{M_f}^{M_i}\frac{\dd M}{\mathcal P(M,a(M))}. \label{eq:geometrical-reciprocal-integral} \end{equation}
They have dimensions of time and describe those parameter paths, with no evaporation interpretation. At fixed $a$,
\begin{equation} \mathcal I_{i\to f}^{(\met,a)}=\frac{k^2}{\gamma}\left[\frac{M_i^3-M_f^3}{3}+\frac{2k+3}{4}a^2(M_i-M_f)\right]+\order(a^4). \label{eq:metric-lifetime-fixed-a} \end{equation}
For fixed $\chi$, define
\begin{equation} x_\ell=\sqrt{1-k\chi^2}, \qquad \mathcal D_\ell=(1+x_\ell)^2+\chi^2. \label{eq:metric-xD} \end{equation}
Thereby,
\begin{equation} 
\mathcal P_{\met}^{(\chi)}=\frac{g_\star\epsilon_{\mathrm{em}}}{480\pi k^2M^2}\frac{x_\ell^4}{\mathcal D_\ell^3}. \label{eq:metric-luminosity-chi} \end{equation}
\begin{equation} \mathcal I_{i\to f}^{(\met,\chi)}=\frac{160\pi k^2}{g_\star\epsilon_{\mathrm{em}}}\frac{\mathcal D_\ell^3}{x_\ell^4}(M_i^3-M_f^3). \label{eq:metric-lifetime-chi} \end{equation}
For the \textit{metric--affine} functional along the same constant $\chi$ path,
\begin{equation} \mathcal I_{i\to f}^{(\MA,\mathrm{loc})}=\frac{M_i^3-M_f^3}{3\gamma}\left[1+\frac54\chi^2+\frac74X+\frac{95}{48}X\chi^2\right]+\order(\chi^4,X^2,X\chi^4). \label{eq:ma-local-lifetime} \end{equation}

\begin{figure}
    \centering
    \includegraphics[scale=0.55]{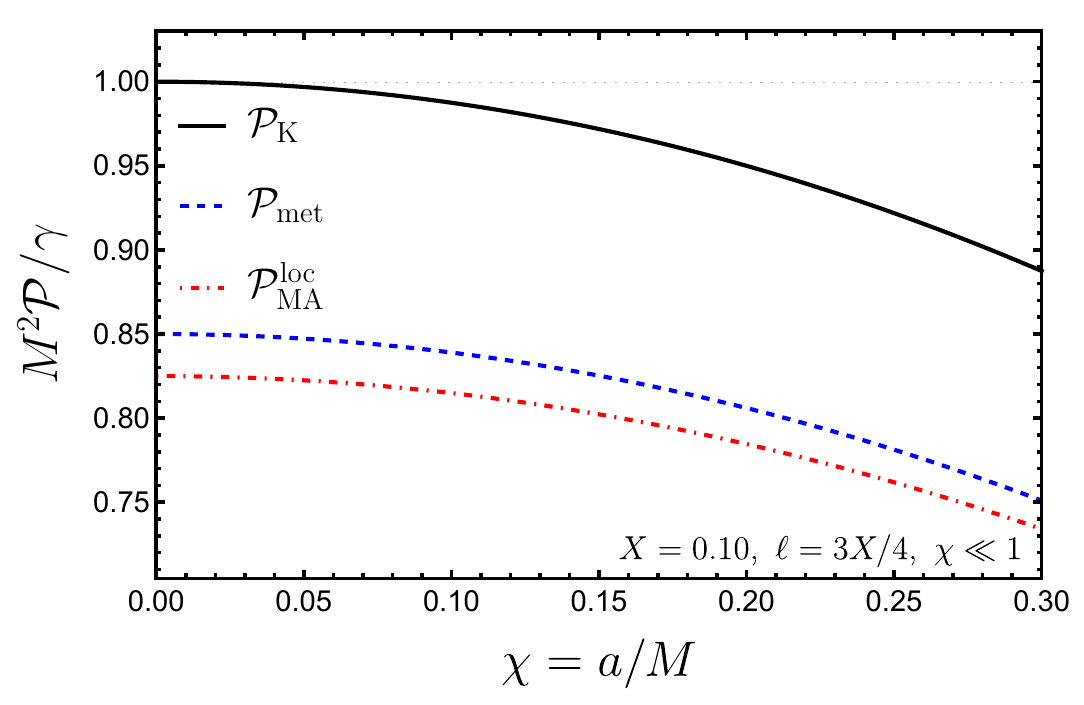}
    \caption{{Dimensionless geometrical functionals $M^2\mathcal P/\gamma$ as functions of the parameter ratio $\chi=a/M$, for $X=0.10$ and $\ell=3X/4$, at identical metric parameters within the weak deformation and slow rotation expansions. }}
    \label{f3}
\end{figure}

\begin{figure}
    \centering
    \includegraphics[scale=0.55]{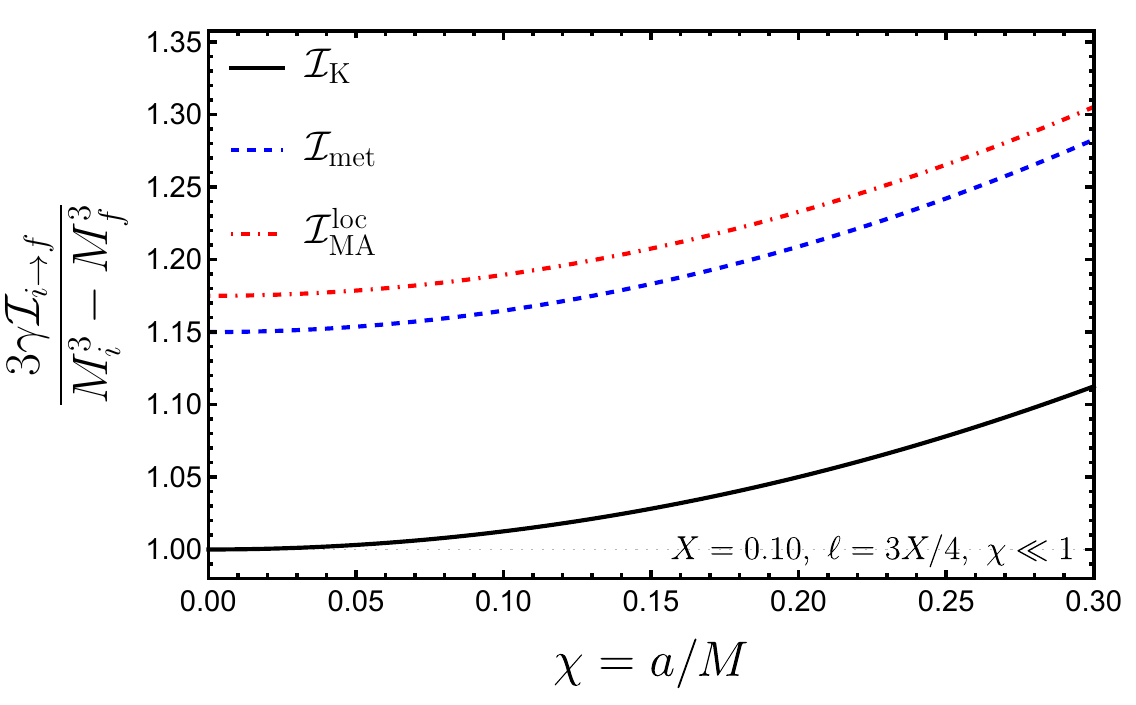}
    \caption{{Normalized geometrical parameter integrals $3\gamma\mathcal I_{i\to f}/(M_i^3-M_f^3)$ as functions of $\chi=a/M$, for $X=0.10$ and $\ell=3X/4$, along constant $\chi$ paths. }}
    \label{f4}
\end{figure}
}

\clearpage
\bibliographystyle{ieeetr}
\let\PointSevenBibitem\bibitem
\renewcommand{\bibitem}[1]{\ifstrequal{#1}{GuoFan2026}{}{\ifstrequal{#1}{Zhu:2026vae}{}{\ifstrequal{#1}{DLMFSpheroidal}{}{\ifstrequal{#1}{Kato1995}{}{\color{black}}}}}\PointSevenBibitem{#1}}
\bibliography{main}
\let\bibitem\PointSevenBibitem
\color{black}

\end{document}